\documentclass[final,hidelinks]{elsarticle}

\usepackage{lineno,hyperref}
\modulolinenumbers[5]
\usepackage{siunitx}
\usepackage{amssymb,amsmath,multicol}
\usepackage{xcolor}
\usepackage{graphicx}
\usepackage{subcaption}
\def\permille{\ensuremath{{}^\text{o}\mkern-5mu/\mkern-3mu_\text{oo}}}

\journal{Measurement}

\begin{document}

\begin{frontmatter}

\title{Pneumatic System for Pressure Probe Measurements in Transient Flows of Non-Ideal Vapors Subject to Line Condensation}

\author[mymainaddress]{Camilla C. Conti\corref{mycorrespondingauthor}}
\ead{camillacecilia.conti@polimi.it}
\author[mysecondaryaddress]{Alberto Fusetti}
\author[mysecondaryaddress]{Andrea Spinelli}
\author[mysecondaryaddress]{Paolo Gaetani}
\author[mymainaddress]{Alberto Guardone}

\address[mymainaddress]{Politecnico di Milano, Department of Aerospace Science and Technology, via La Masa 34, 20156, Milano, Italy}
\address[mysecondaryaddress]{Politecnico di Milano, Energy Department, via Lambruschini 4, 20156, Milano, Italy}

\cortext[mycorrespondingauthor] {Corresponding author}

\begin{abstract}
This paper presents the design, construction and commissioning of a pneumatic system for pressure probe measurements in 
flows of organic vapors in non-ideal conditions, namely in the thermodynamic region close to the liquid-vapor saturation curve and the critical point where the ideal gas law is not applicable. \\
Experiments were carried out with fluid siloxane MM (hexamethyldisiloxane, C$_6$H$_{18}$OSi$_2$), commonly employed in medium/high temperature Organic Rankine Cycles (ORCs), in the Test Rig for Organic VApors (TROVA), a blow-down wind tunnel at the Laboratory of Compressible fluid dynamics for Renewable Energy Applications (CREA lab) of Politecnico di Milano. TROVA operation is intrinsically transient due to its batch nature, with a low frequency content ($\sim \SI{1}{Hz}$) related to the emptying of the high-pressure reservoir feeding the test section. \\
The challenges linked to possible condensation in pneumatic lines, such as vapor-liquid menisci, hydrostatic head and mass-sink effects, were evaluated by means of theoretical calculation and experiments. To avoid these issues, a nitrogen-flushed pneumatic system for absolute and differential pressure measurements was designed and successfully tested with superheated MM vapor expanding in planar choked converging nozzles characterized by a portion with constant cross-sectional area yielding design Mach numbers of $0.2$, $0.5$ and $0.7$.
Commissioning of the complete system including a probe was performed with the first ever testing of an L-shaped Pitot tube in non-ideal subsonic flows of siloxane MM vapor at Mach numbers $M=0.2$ and $0.5$. Measurement delay issues were identified and assessed through a dynamic testing procedure, and were solved by reducing the overall pneumatic lines volume including the one hidden within pressure transducers.
The correct performance of the complete system was therefore verified for probe measurements of total, static and dynamic pressure in non-ideal flows of organic vapors. 
This sets the foundation for future directional pressure probes calibration and use in the characterization of such flows, as in direct measurement of total pressure losses across shocks and in testing of ORC turbine blade cascades.

\end{abstract}


\end{frontmatter}

\nolinenumbers

\section{Introduction}

Non-ideal compressible fluid dynamics is a branch of gas dynamics involving flows whose behaviour strongly differs from the ideal gas one due to operation in thermodynamic regions close to the liquid-vapor saturation curve and the critical point where intermolecular forces are not negligible \citep{Thompson1971}. 
An indication of the level of non-ideality is notably given by the value of the \textit{compressibility factor} $Z = \dfrac{P}{R T \rho}$, where $P$ is pressure, $T$ is temperature, $\rho$ is density and $R$ is the gas constant.
$Z$ is identically equal to $1$ in case of ideal gas behaviour and possibly differs from $1$ otherwise, as it represents the departure of the fluid volumetric behavior from that of an ideal gas at same temperature and pressure.

Non-ideal flows occur in a vast range of engineering processes, from rocket propulsion to industrial and chemical activities. Examples are the oil $\&$ gas, heat pumps and refrigeration fields, and even pharmaceuticals production with the use of the rapid expansion of supercritical solutions \citep{Helfgen2003}.
In the power generation field, non-ideal flows are found in supercritical carbon dioxide (\textit{sCO$_2$}) power cycles and in Organic Rankine Cycles (\textit{ORCs}).
The latter are a convenient option for low/medium size power generation from renewable energy and waste heat recovery sources at low/medium temperatures, thanks to their low cost, plant simplicity and thermodynamic efficiency \citep{Colonna2015,Macchi2016}. 
ORCs involve flows of organic compounds with high molecular mass and complexity in the close proximity of the saturation curve. As a result, turbine flows differ from standard turbomachinery ones because they are highly supersonic and show significant non-ideal gas effects, such as flow field dependence on stagnation conditions, possible increase of the speed of sound and non-monotonic trend of the Mach number along an expansion \citep{Colonna2006,Spinelli2019}. 
Due to the peculiarity of non-ideal vapor flows in ORCs, measurements such as velocity, mass flow rate or turbine performance, which are routinely carried out in standard cycles and turbomachinery (e.g. gas turbines operating with air and combustion gases), are not commonly performed yet.
An important issue in real ORC operating plants is indeed the closure of mass and energy balances due to the lack of reliable mass flow rate measures \citep{Zanellato2017}. Even blade cascade testing, quite common in the design process of gas and steam turbines, is only starting to take place for such flows \citep{Baumgartner2019,Baumgartner2020}.\\
All of this is mainly because no appropriately calibrated instrumentation for non-ideal conditions is available since there is a lack of ad-hoc calibration facilities. The reason for the latter is two-fold. 
First of all, ORC working fluids are liquids at standard room temperature and pressure. Typical inlet turbine flows are  instead at saturated, superheated or supercritical conditions, with temperatures and pressures ranging from about $100$ to $\SI{400}{\celsius}$ and $10$ to $\SI{50}{bar}$ \citep{Macchi2016}. Thus, in order to reproduce realistic conditions in a wind tunnel for testing and calibration purposes, a closed gas cycle or a phase transition thermodynamic cycle must be put in place. These are noticeably more complicated and expensive with respect to operation with incondensable gases such as air, where compressed air storage tanks or continuous loops are sufficient to carry out a calibration campaign. Moreover, measurement procedures are also more complex due to the high fluid temperature involved and condensation issues in pneumatic lines.
The second reason is linked to non-ideal gas effects and, in particular, to flow field dependence on stagnation conditions. If isentropic expansion of an ideal gas in a fixed geometry is considered, trends in Mach number, pressure and temperature ratios are independent of the actual total conditions, and only depend on molecular complexity through the specific heats ratio $\gamma$. 
This means that calibration flows need not cover the whole range of possible total temperature and pressure conditions in which a probe might operate unless Reynolds number effects on the probe head are significant.
If non-ideal flows are considered instead, flow behaviour is dependent on both the fluid and stagnation conditions. This complicates the calibration procedure and limits its generality, making it very case-dependant and increasing its cost considerably, since flows at all possible total operating conditions need to be reproduced. \\
For all these reasons, there are currently very few wind tunnels for organic vapors in the world and none of them are yet used as probe calibration facilities for non-ideal flows.
Amongst active \textit{nozzle-fitted} facilities (those with test sections featuring simple geometries such as nozzles) suitable for pressure probes testing, is the Test Rig for Organic VApors (TROVA) \citep{Spinelli2013} at the Laboratory of Compressible Fluid-dynamics for Renewable Energy Applications (CREA Lab) of Politecnico di Milano, where all the experimental campaigns concerning the present work were carried out. Other plants of this kind are the ORCHID at TU Delft \citep{Head2016}, the CLOWT at Muenster University of Applied Sciences \citep{Reinker2017} and the dense-gas blowdown facility at Imperial College London \citep{Robertson2019}.
Several \textit{turbine-fitted} facilities such as the LUT micro-ORC test rig at Lappeenranta – Lahti University of Technology \citep{Turunen-Saaresti2016} exist too. The ORCHID at TU Delft is designed to also operate in this configuration. These types of plants are mainly devoted to performance measurement of the different components and of the overall thermodynamic cycle and are thus more suitable for pressure probes employment as measurement instruments rather than for their calibration. \\
Research efforts are now starting to move towards the use of pressure probes in non-ideal flows. The first published works on the topic are from the CLOWT plant at Muenster University of Applied Sciences. Results on the performance of a rotatable cylinder Pitot probe in high subsonic flows with fluid NovecTM 649 were recently presented \citep{Reinker2020} as part of a preliminary study to establish measurement techniques for the determination of Mach numbers in high-subsonic and transonic organic vapor flow fields.

This paper presents the development, testing and commissioning of a pneumatic system for pressure measurements that would enable the calibration of probes operating with non-ideal flows representative of ORC working conditions. 
The procedure is illustrated step by step, from the choice of particular nozzle geometry and subsonic operating conditions, to initial testing with no probes in absolute and differential pressure measurement configurations, until the final commissioning with a Pitot tube in subsonic conditions, also considering dynamic response issues. 
Such a detailed account can be useful for anyone wishing to develop a pneumatic system for pressure probes testing in similar conditions, even beyond the ORC field. \\
The pneumatic system is composed of lines and cavities of different shapes and sizes connecting pressure taps at the measuring point on the probe or at the test section wall with pressure transducers mounted at line ending. It is implemented in the \textit{Test Rig for Organic VApors} (TROVA), where testing with fluid siloxane MM (hexamethyldisiloxane, C$_6$H$_{18}$OSi$_2$) was carried out. An outline of the experimental apparatus is presented in Section \ref{sec_EXPAPP}.
Pneumatic lines are subject to condensation unless line heating is supplied. This can lead to poor pressure measurements quality related to presence of vapor-liquid menisci, hydrostatic head and mass-sink effects. These challenges were evaluated by means of theoretical calculation and experiments in the context of TROVA operation and are detailed in Section \ref{sec_PLnoN2}. 
A pneumatic system configuration involving nitrogen flushing of the line was found to be the optimal setup to overcome all aforementioned difficulties, as reported in Section \ref{sec_PLflushN2}. 
This scheme was applied to differential measurements for probes testing, and its specific layout for experimentation with Pitot tubes is found in Section \ref{sec-LinesPitotSub}. Its commissioning during the first ever experimental campaign involving an L-shaped Pitot tube in non-ideal subsonic flows of siloxane MM vapor is also presented in Section \ref{sec-PitotMMsub}, together with the identification of a measurement delay issue.
Analysis and further testing on the latter are recounted in Section \ref{sec-pnemDynTest}, together with final results verifying the correct functioning of the improved pneumatic system configuration. 
Section \ref{sec-conclusions} finally draws the conclusions of this work. \\

\section{Experimental Setup}
\label{sec_EXPAPP}
\subsection{Test Rig for Organic VApors - TROVA}
The \textit{Test Rig for Organic VApors} (TROVA) is a blow-down wind tunnel built with the aim of characterizing non-ideal flows of organic vapors, especially those representative of turbine expansions in ORCs. Different fluids can be employed and experiments reported here were carried out with fluid siloxane MM (hexamethyldisiloxane, C$_6$H$_{18}$OSi$_2$), commonly employed in medium/high temperature ORCs.\\
The working fluid is isochorically heated in a High Pressure Vessel (HPV) until desired temperature and pressure are reached. It is then discharged to a Low Pressure Vessel (LPV) by passing through a settling chamber (plenum) and expanding in the test section through a purposely designed planar nozzle. \\
TROVA operation is intrinsically transient due to its batch nature.
Before the beginning of a test, the test section and LPV are vacuumized to MM saturation pressure at room temperature ($P\sim \SI{50}{mbar}$).
After test start ($t_{start}$), a peak is reached and then pressure in the test section decreases in time with a low frequency content ($\sim \SI{1}{Hz}$) related to the emptying of the HPV \citep{Spinelli2018}. \\
Due to the decreasing pressure linked to the plant batch nature, the most non-ideal flow conditions are achieved at the beginning of each test.\\
A more detailed description of the plant and its design can be found in \citep{Spinelli2013}.\\

\subsection{Nozzle Expansion Characterization and Instrumentation}
\label{sec-instruments}
\begin{figure}
	\begin{center}
		\includegraphics[width=1\textwidth,trim={0 27cm 0 15cm}, clip]{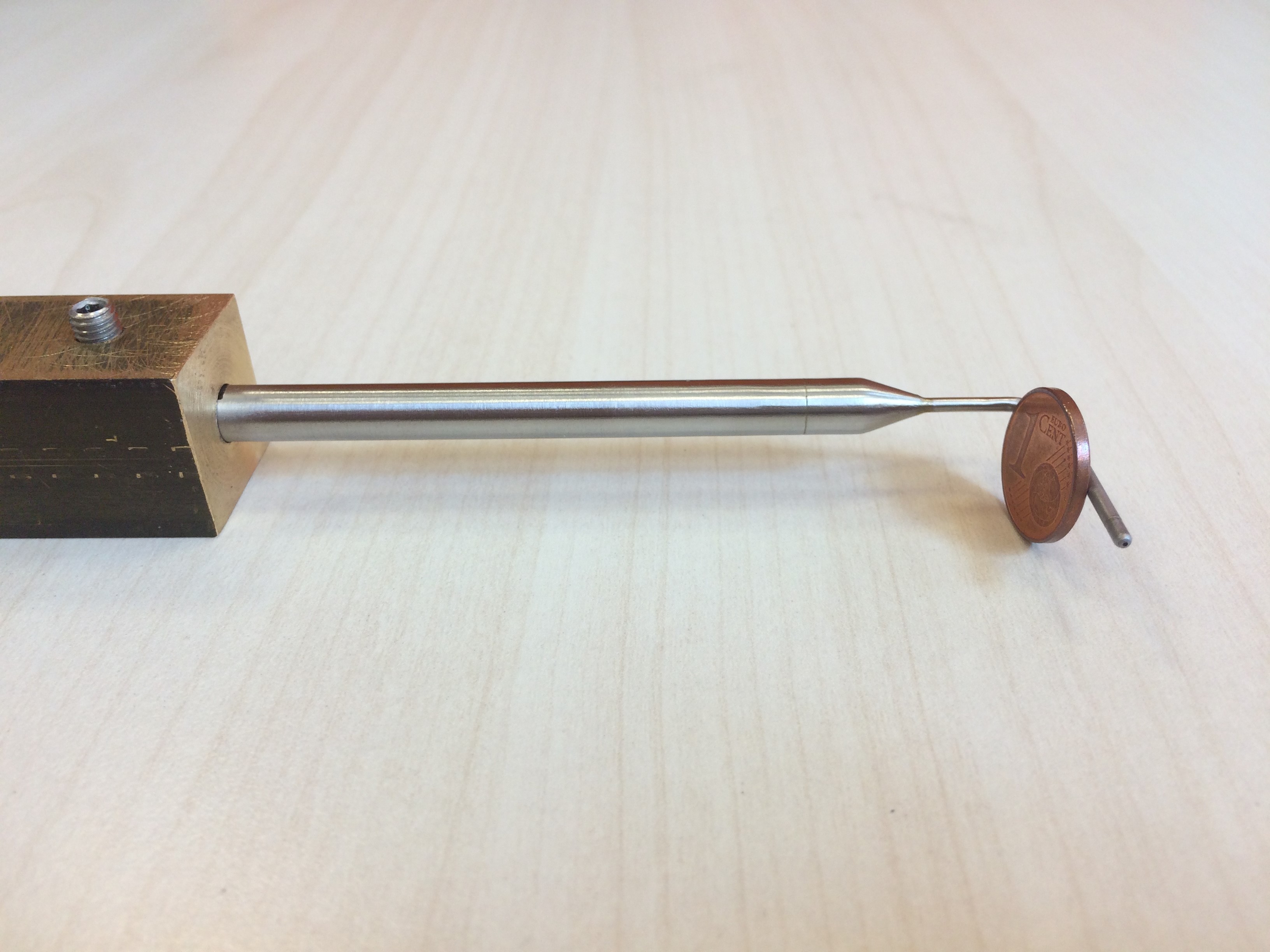}
		\caption{L-shaped total-static Pitot tube employed for pneumatic system commissioning. } 
		\label{fig-PitotAP}
	\end{center}
\end{figure}

Flow expansions are characterized by total conditions measurements in the plenum upstream of the test section, where velocity is low enough (about $\SI{1}{m/s}$) for kinetic energy to be negligible. Thus, total pressure $P_T$ is measured at a wall tap with an absolute pressure transducer and total temperature $T_T$ with thermocouples. Total temperature and pressure can vary in the range $200 - \SI{260}{\celsius}$ and $7 - \SI{24 }{\bar}$ respectively for fluid MM.
Flow along the nozzle axis is characterized by static pressure measurements performed by a transducer mounted on the test section rear plate, connected through a \SI{30}{mm} long line-cavity system to machined wall taps of $\SI{0.3}{mm}$ in diameter. 
The frequency response of this line-cavity system was estimated well above that required to capture the pressure level changes linked to the emptying of the HPV \citep{Spinelli2018}.\\
Back-plate mounted absolute piezoresistive transducers are exposed to high temperature organic vapor flows. Thus, due to their temperature sensitivity, transducers must be calibrated both in pressure and temperature from vacuum to full scale ($\SI{3.5}{bar} \leq FS \leq\SI{40}{bar}$) in the range $25 - \SI{300}{\celsius}$. Thermocouples are calibrated in the same temperature range.
The expanded uncertainty is $0.07$~\% of the full scale for pressure sensors and about $\SI{1}{\celsius}$ for thermocouples. Data sampling frequency for temperature and pressure measurements is $\SI{1}{kHz}$ and signals are averaged over $100$ time instants yielding a time resolution of $\SI{0.1}{s}$.
Further details on \textit{TROVA} instrumentation can be found in \citep{Spinelli2018}. \\
The L-shaped total-static Pitot tube in Figure \ref{fig-PitotAP}, with a stem length of $\SI{25}{mm}$, outer diameter of $\SI{1.6}{mm}$, total pressure tap diameter of $\SI{0.6}{mm}$ and six taps on the static ring, was employed for the commissioning phase of the complete pneumatic system. 

\subsection{Employed Nozzles and Rear Plates}
\begin{figure}
	\begin{center}
		\begin{subfigure}[t]{1\textwidth}
			\centering	
			\includegraphics[width=1\textwidth,trim={0 0 0 0}, clip]{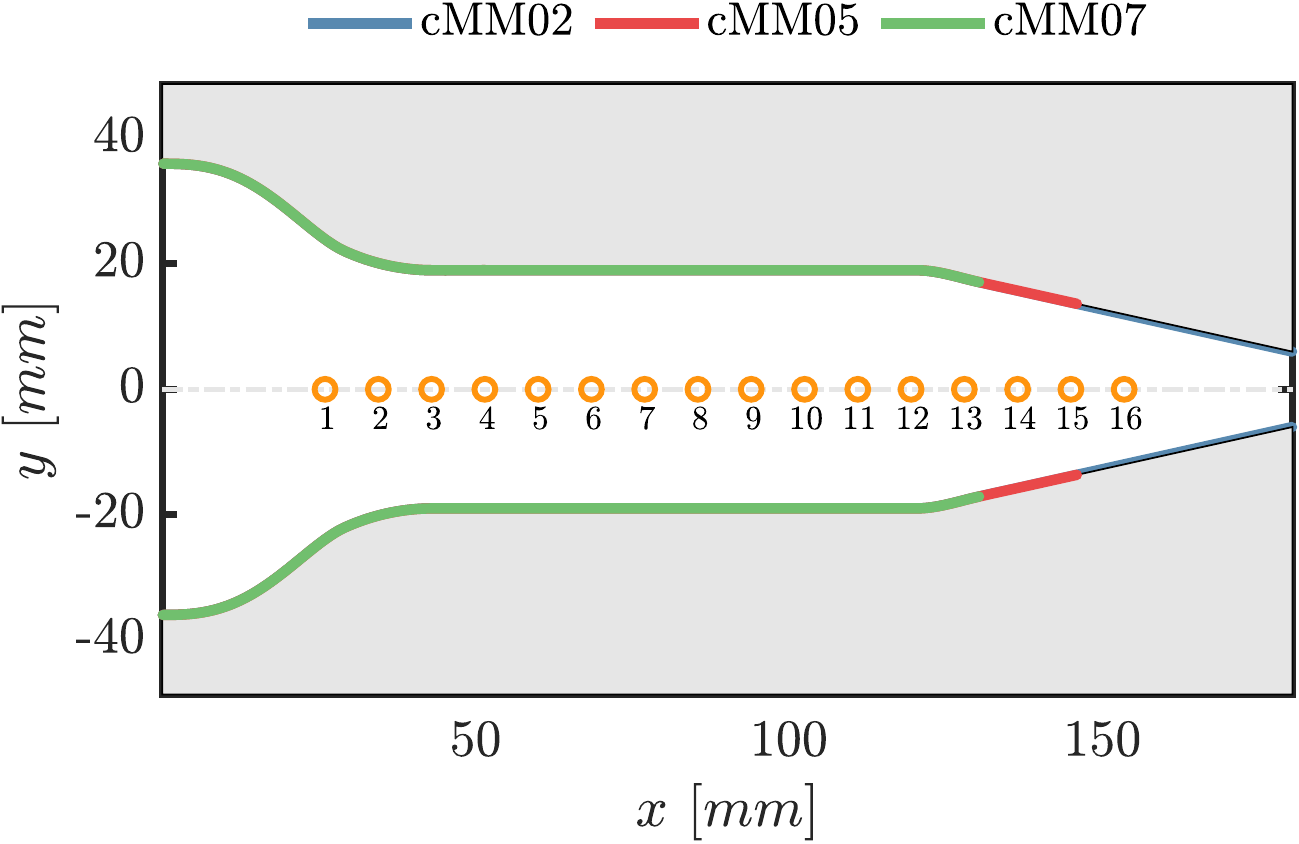}
			\caption{Configuration with \textit{HSR} rear plate. Pressure taps are also shown.  }
			\label{fig-convJLo}
		\end{subfigure}
		\vspace{2pt}	
		
		\begin{subfigure}[t]{1\textwidth}
			\centering	
			\includegraphics[width=1\textwidth,trim={0 0 0 0}, clip]{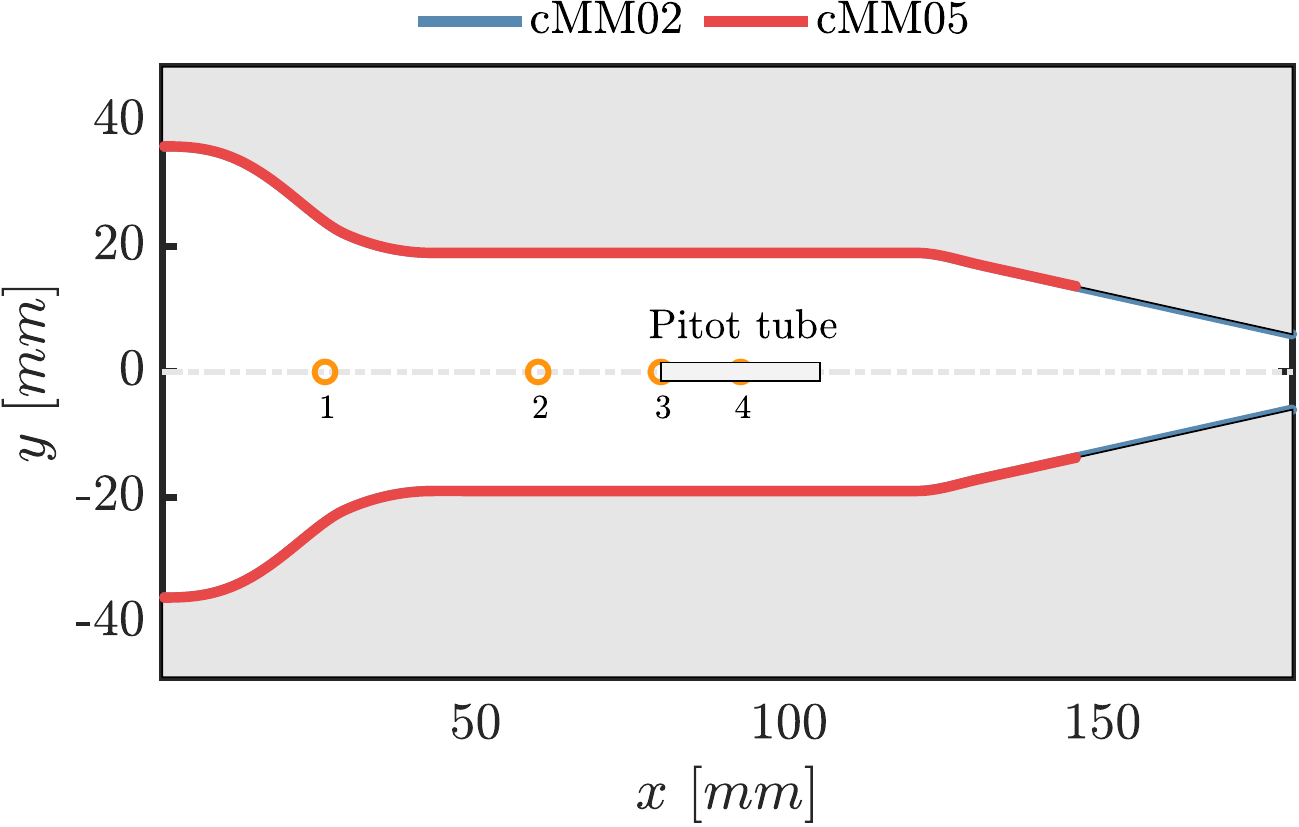}
			\caption{Configuration with \textit{Pitot} rear plate and Pitot tube. Pressure taps are also shown.}
			\label{fig-convPitot}
		\end{subfigure}	
	\end{center}
	\caption{Choked converging nozzles for three different Mach numbers $M=0.2,0.5,0.7$ in the constant cross section region. } 
	\label{fig-convergenti}
\end{figure}

Superheated MM vapour expands in planar choked converging nozzles characterized by a portion with constant cross-sectional area giving design Mach numbers of $0.2$, $0.5$ and $0.7$. As represented in Figure \ref{fig-convergenti}, nozzles are named accordingly as $cMM02$, $cMM05$ and $cMM07$.\\
The first convergent section of the nozzle is designed using a 5th order polynomial, yielding a double concavity that provides gentle flow acceleration up to the design Mach number and reduces flow disuniformities. 
This portion and the constant cross-section one (semi-height $h=\SI{19}{mm}$) are the same for all nozzles at different Mach numbers. 
The second convergent is a straight line that ends with the throat, which is always choked due to the very low pressure in the LPV. The slope is the same for all Mach numbers but the length is such that the area ratio $A^*/A$ between throat and constant cross-section corresponds to the desired Mach number at total design conditions $P_T=\SI{5}{bar}$ and $T_T=\SI{210}{\celsius}$. Due to the aforementioned dependence on stagnation conditions, the Mach number varies slightly  during tests as $P_T,T_T$ change following the characteristic transient of the plant.\\
Nozzle depth is imposed by the test section to $\SI{18.7}{mm}$ and geometrical dimensions are reported in Table \ref{tab-NozzleGeometry}.\\
The choice of this particular design was deliberate, because probe insertion in the constant cross-section portion will allow future calibration of pressure probes with flows of organic vapours at a certain subsonic Mach number but varying levels of non-ideality. 
During preliminary testing of the pneumatic system, this nozzle design allows instead the evaluation of system performance before probe insertion.
To this purpose, nozzles are mounted on the \textit{HSR} (High Space Resolution) rear plate featuring $16$ wall taps, $\SI{8.5}{mm}$ apart, as shown in Figure \ref{fig-convJLo}.
This configuration enables the direct comparison of pressure measures at the end of the pneumatic line connected to a certain wall tap with transducers mounted at adjacent ones. Indeed, compared measurement points are all close to one another and located in the constant cross-section portion of the nozzle, so the same pressure should be found when probe and boundary layer blockage are negligible. \\
For the commissioning of the complete pneumatic system, the Pitot tube was inserted in the constant cross-section portion of nozzles \textit{cMM02} and \textit{cMM05}.
Nozzle \textit{cMM07} was not considered in this phase because 1D calculation indicated that the probe presence in the test section would cause non-negligible blockage effects with considerable flow acceleration possibly up to transonic flow conditions.\\
Commissioning tests were carried out using the \textit{Pitot} rear plate illustrated in Figure \ref{fig-convPitot}. 
It features four static pressure taps and a hole with a diameter large enough to fit the Pitot tube stem. 
The hole is located on the nozzle axis in the constant cross section region and sufficiently distant from both the first and second convergents to ensure that the probe tip lays in a position where flow properties are uniform, as confirmed by numerical simulations performed during the design phase of the experiment and not detailed here for brevity. \\
Taps $2,3$ and $4$ are all located in the constant cross section region of the employed nozzles.
Tap $3$ is in correspondence of the Pitot tube total pressure hole and tap $4$ is in correspondence of its static pressure ring. \\
This configuration allows to evaluate the performance of the complete pneumatic system (probe included) through direct comparison with plant reference counterparts. 
Since the flow is subsonic, the total pressure measured by the Pitot tube can be compared with the reference total pressure in the plenum. Pitot tube static pressure can instead be evaluated against the one at the other wall taps in the constant-section region. \\

\begin{table}[]
	\centering
	\begin{tabular}{ccccccccc}
		\hline
		Name                 & $H_{th}$  & $H_{in}$ & $H_{out}$ & $x_{out}$  & $P_t$    &  $T_t$         &  $M$     \\
		& $[mm]$    & $[mm]$   & $[mm]$    &    $[mm]$  & $[bar]$  &  $[^{\circ}C]$  &  $[-]$     \\
		\hline
		\textit{cMM02}       & 6.0       & 36	     & 6.0	    & 180.4	     &  5	    &210	         &0.2    \\
		\textit{cMM05}       & 13.6      & 36	     & 13.6	    & 146.0	     &  5	    &210	         &0.5    \\
		\textit{cMM07}       & 17.1      & 36	     & 17.1	    & 130.4	     &  5	    &210	         &0.7    \\	
		\hline  
		
	\end{tabular}
	\caption[]{Nozzle semi-height at the throat, inlet and outlet, and x-coordinate of the outlet section together with design total conditions and Mach number in the constant cross-section region. }
	\label{tab-NozzleGeometry}
\end{table}

\subsection{Choice of Experimental Configuration}

Concerning the overall pneumatic system commissioning, the choices of probe type, subsonic conditions, nozzle shape and pressure tap positions were deliberate. It is only their combined testing that allows to directly evaluate the performance of the complete pneumatic system against reference quantities and check the proper functioning of the dedicated lines. 
Indeed, a Pitot tube is the only type of probe that, with no calibration and if aligned with the flow in subsonic conditions, measures both total and static quantities that have a direct reference counterpart. 
Had directional pressure probes been used, no direct reference would have been available for static pressure measures. Indeed, these types of probe often have a total pressure tap but do not have a static tap parallel to flow direction whose measure can be compared with reference readings from the wind tunnel. 
Analogously and as previously anticipated, the particular convergent nozzle shape with a constant cross section (with several static pressure taps), is extremely important to provide a direct static pressure reference and allow the clear identification of any possible issues with the dedicated pneumatic line.
Moreover, had supersonic flows been directly considered, total pressure shock losses would not have permitted any direct reference, but comparison would have required additional shock losses calculation. \\
Thus, the specific experimental configuration tested here is a key preparatory step for future use and calibration of any kind of probe, such as directional pressure probes, in both subsonic and supersonic conditions. \\

\subsection{Unheated Pneumatic Lines}
\begin{figure}
	\begin{center}
		\includegraphics[clip, trim=0.5cm 0cm 10cm 2.5cm, width=1\textwidth]{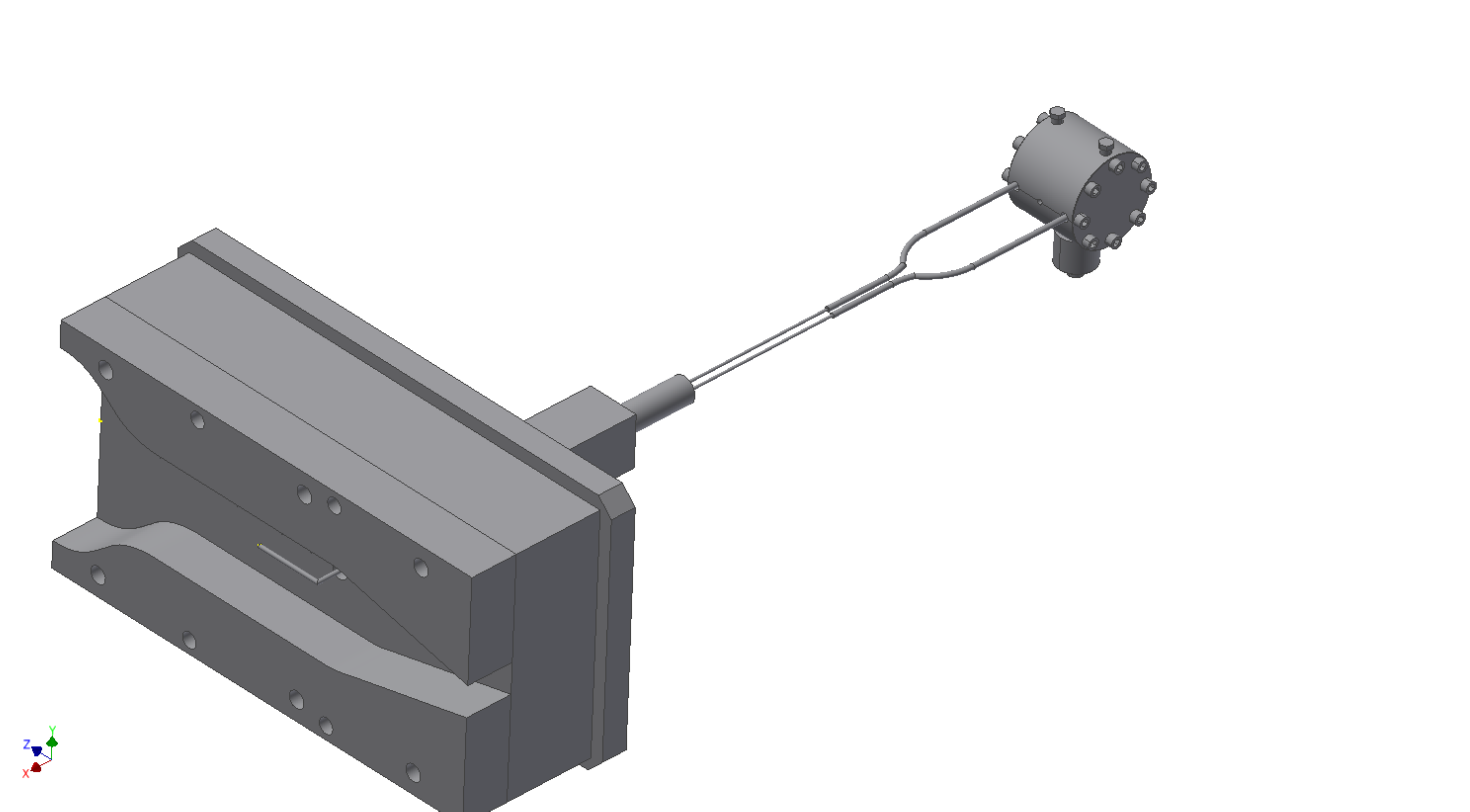}
		\caption{Differential pneumatic lines for pressure probe measurements. In this example, Pitot tube kinetic head is being measured.}
		\label{fig-pneumSys}
	\end{center}
\end{figure}

All pressure measures in the TROVA to date were performed with the use of absolute transducers directly mounted on the back plate of the heated test section. This solution ensures response times well below the characteristic HPV emptying time and avoids possible condensation issues in pneumatic lines. The downsides are more expensive transducers due to high temperature operation and time-consuming calibration in temperature as well as pressure. 
However, if probes are employed, pneumatic lines are inevitably present and differential pressure measures are carried out between the various probe taps so as to minimize the final measurement uncertainty. Also, in case Pitot tubes are tested, the difference between reference TROVA pressure and the one at the probe is also to be acquired (for both total and static quantities).
A differential configuration is not possible with plate-mounted transducers: lines connected to wall pressure taps, exiting the back of the test section and meeting at the two ends of a differential transducer are needed, as shown in Figure \ref{fig-pneumSys}. 
The length of these lines can vary depending on the distance between measurement points.

Since fluid MM is liquid at room temperature and considered operating pressures during tests, pneumatic lines are subject to condensation unless heating is supplied. The latter solution is complicated due to the small line diameter necessary for reduced response times ($\leq \SI{2}{mm}$ internal, $3-\SI{4}{mm}$ external), is expensive, cumbersome and not flexible in case line length is varied. Moreover, if the line is unheated, the fluid at the transducer sensor will be significantly cooler than at the measurement point. This is positive, because differential transducers able to withstand high temperatures are not readily available on the market. Also, temperature calibration is no longer required. For these reasons, line heating was not pursued. Rather, the focus was on overcoming practical challenges linked to line condensation, as explained in the following. 

\section{Pneumatic Lines with Condensation}
\label{sec_PLnoN2}

Initial testing was carried out on a single pneumatic line of $\SI{2}{mm}$ internal diameter and $\SI{200}{mm}$ long. This length  is the estimated minimum one for probe testing. \\
To verify that condensation does take place and determine the possible fluid temperature at which sensors might be exposed, a thermocouple was first placed at the end of the line. The latter was connected to the wall tap in the plenum where total pressure is measured because this is where maximum temperature occurs. The peak temperature read by the thermocouple for all tests was less than $\SI{40}{\celsius}$, confirming that condensation occurs as the fluid enters the line and that sensors can expect a fluid temperature not far from ambient one, considering that lines will likely be longer in actual probe testing configuration. \\
Condensation can be problematic, even during steady measures, for two reasons: menisci presence and hydrostatic head.
The presence of vapour and liquid phases in the line implies the existence of menisci and a consequent pressure difference across the interface which can alter the reading at the sensor at the end of the line. This pressure difference was estimated using the Laplace-Young law and experimentally verified values of surface tension and contact angle for fluid MM with a stainless steel surface ($\sim \SI{0}{^\circ}$).
The calculated value was $\SI{32}{Pa}$, largely below transducers uncertainty. 
Moreover, since it is impossible to predict where in the line condensation starts, liquid formation could create a hydrostatic head between the vapour-liquid interface and the transducer location. 
This effect is considered negligible if the pressure difference caused by the hydrostatic head is lower than the transducer uncertainty. 
For instance in case of a \SI{500}{Pa} uncertainty (typical of a \SI{5}{bar} full scale piezoresistive transducer), calculation for liquid MM showed that this effect is indeed negligible if the geodetic head is lower than $\SI{70}{mm}$.
Particular care was thus taken to ensure this during line installation.

For all further testing, the thermocouple at the end of the line is replaced by an absolute transducer. 
Signal from the latter is compared with a plate-mounted transducer at another identical wall pressure tap in the plenum. The line transducer is expected to correctly measure flow pressure once the whole line volume $V_{line}$, which is vacuumized before each test, is filled with MM. 
The corresponding time delay, defined as the time taken from test start for the line transducer signal to be within error bars of the plate-mounted one (taken as reference), can be estimated as follows. 
Vapour flow through the wall tap to the line is assumed choked at the minimum cross section $A_{min}$. This is a realistic hypothesis at the beginning of a test when lines are under vacuum conditions. The total pressure tap is large, so here $A_{min}$ corresponds to the line internal diameter of \SI{2}{mm}. The time delay $\Delta \tau$ is determined from the balance $V_{line} = \dot{V}_{vap} \Delta \tau = C_d c A_{min} \Delta \tau $ where $\dot{V}_{vap}$ is the volumetric flow rate and $C_d$ is the discharge coefficient taken as $0.65$ for small and sharp edged orifices \citep{Shawney2013}. 
Speed of sound $c$ is calculated from total pressure and total temperature using a Helmoltz energy based fundamental relation of Span-Wagner type embedded in the \textit{RefProp} library \citep{Span2003}.
The calculated time delay of the line was found to be two orders of magnitude below the dataset time resolution of $\SI{0.1}{s}$.

\begin{figure}
	\begin{center}
		\includegraphics[width=0.7\textwidth]{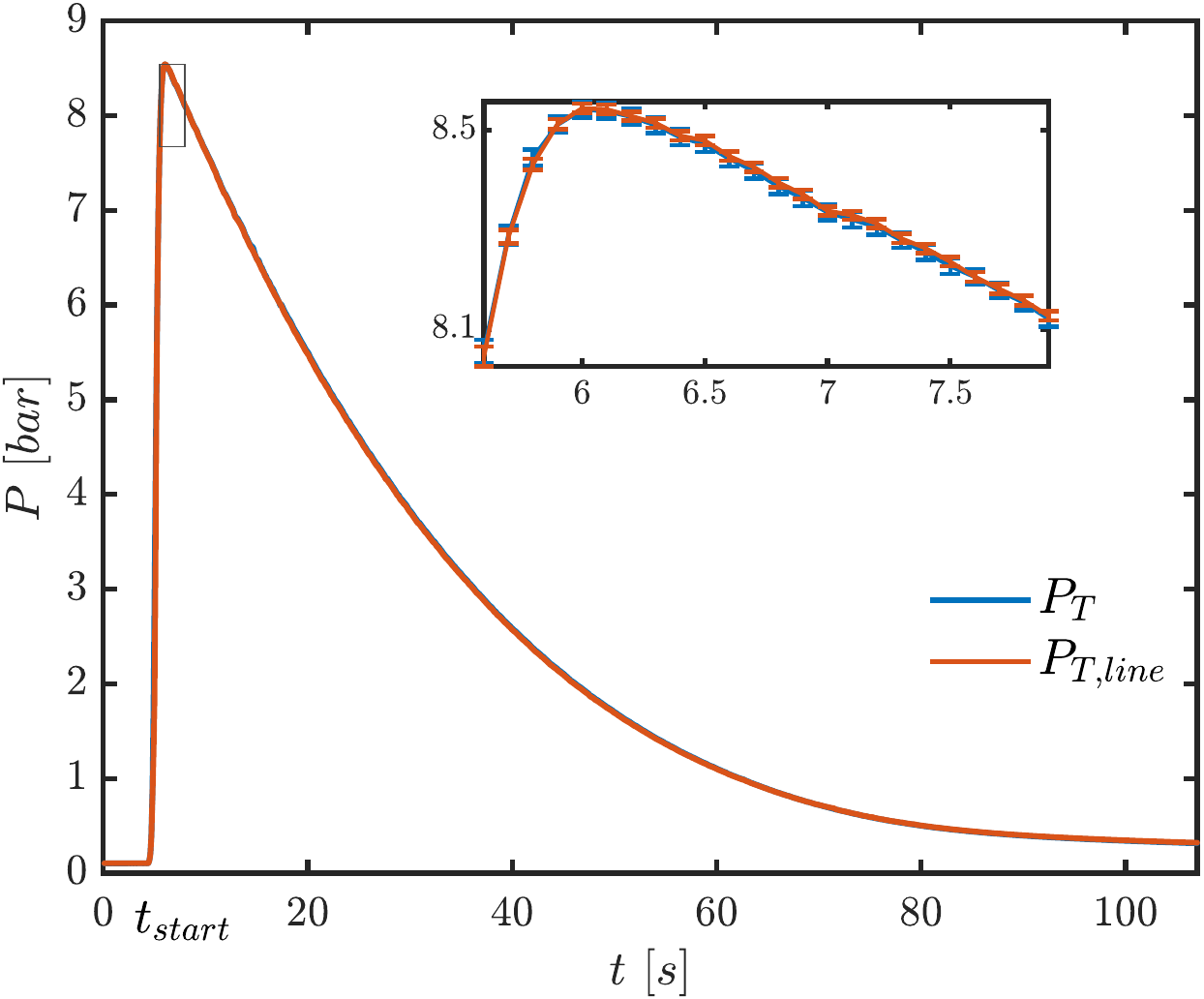}
		\caption{Pressure measures during \textit{Test A}. Error bars are included in the zoom. } 
		\label{fig-PTline}
	\end{center}
\end{figure}

Figure \ref{fig-PTline} shows results for an exemplary test (\textit{Test A}) carried out on nozzle \textit{cMM07} having initial $P_T=\SI{8.5}{bar}$ and $T_T=\SI{215}{\celsius}$ and a significant level of non-ideality with a compressibility factor evaluated at total conditions $Z_T = 0.76$.
Perfect overlap between the two pressure readings was found, without any delay effects due to the line filling and no menisci or hydrostatic head effects. 

\begin{figure}
	\begin{center}
		\includegraphics[width=0.7\textwidth]{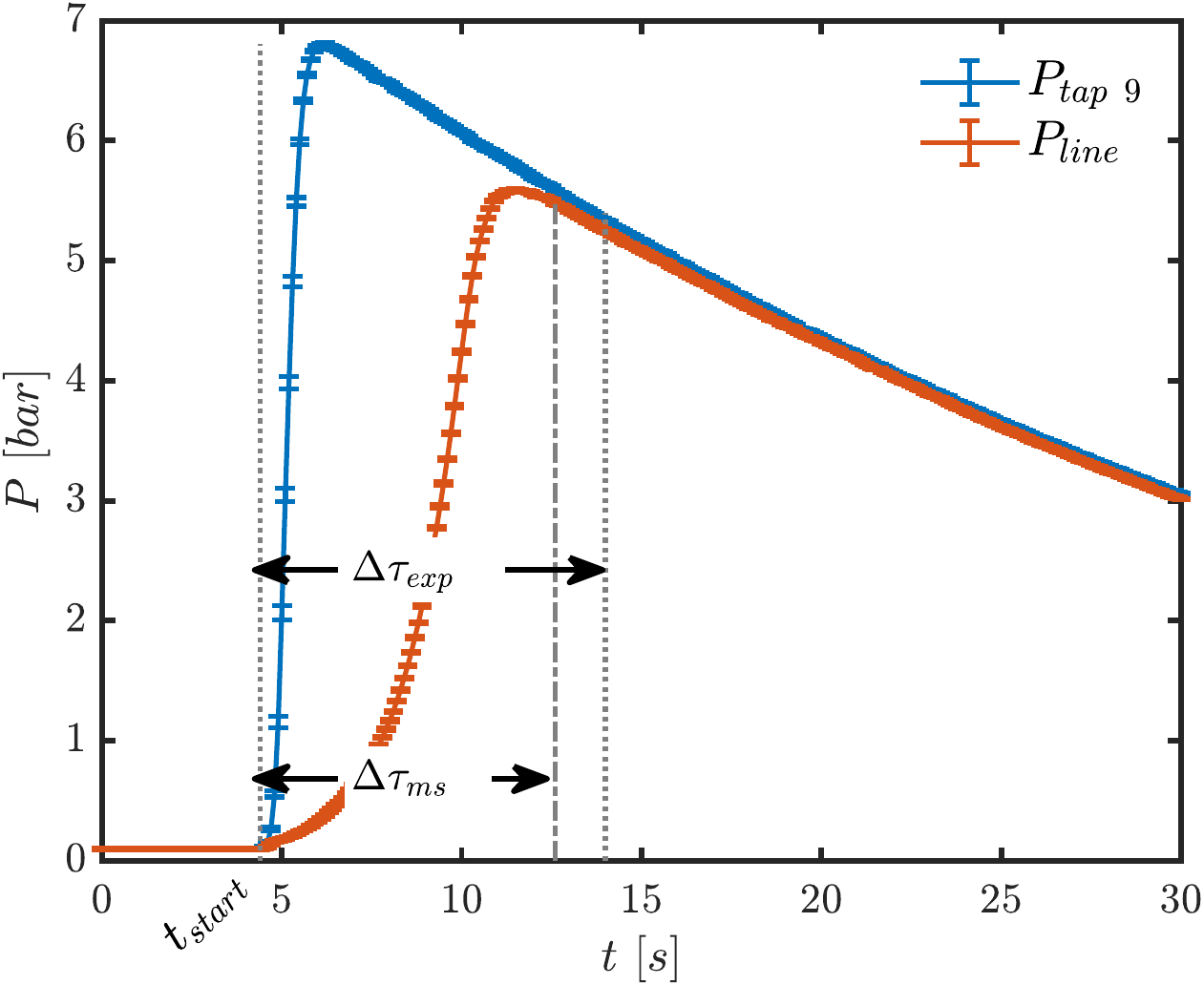}
		\caption{Pressure measures during \textit{Test B}. The first $\SI{30}{s}$ only are shown here for plot clarity. }
		\label{fig-Pline}
	\end{center}
\end{figure}

The pneumatic line was then connected to the static pressure tap number $11$ of rear plate \textit{HSR}, located in the constant area region of the nozzle. The pressure signal was compared with plate-mounted transducers at wall tap $9$ in the same region. 
The time delay of this line was estimated in the same way as \textit{Test A}. Thermodynamic quantities at the static tap were found using total pressure, total temperature and the isentropic flow hypothesis coupled with the aforementioned thermodynamic model. The minimum area in this case is the tap hole of \SI{0.3}{mm} in diameter, leading to larger time delay with respect to the previous configuration, but still well below the dataset time resolution.
As Figure \ref{fig-Pline} shows for an exemplary test (\textit{Test B}) having the same nozzle and initial total conditions as \textit{Test A}, there is instead a significant time delay between signals ($\Delta \tau_{exp} \sim \SI{9.6}{s}$) 
which is orders of magnitude higher than the estimated one. 
As explained in the following section, it is attributable to the \textit{mass sink effect} linked to fluid condensation inside the pneumatic line. 

\subsection{Mass Sink Effect}
\label{sec-masssink}

As the test begins and MM enters the line as vapour, the sensor at its end is unable to correctly measure flow pressure until the whole line volume $V_{line}$ is filled with liquid, given condensation. 
This change of state was neglected in previous delay estimates and is the reason for their disagreement with \textit{Test B} experimental results. Indeed, the higher liquid density means that a larger MM mass will be needed to fill the line although the entering mass flow rate is still limited by the lower vapor density.
This leads to much larger delay times with respect to no line condensation. \\
To verify this \textit{mass sink} phenomenon, a simple model for delay time estimate was developed and verified against experimental data. \\ 
Pressure signal from a plate-mounted transducer is discretized into $z$ time steps of equal size $\Delta t$. 
Thermodynamic properties at each time instant are calculated as described in Section \ref{sec_PLnoN2} and flow through wall tap to the line is again assumed choked at the minimum cross section $A_{min}$.
The vapor mass entering the line at each time step is $dm=d\dot{m} \Delta t = C_d \rho c A_{min} \Delta t$. 
When the total vapor mass $m_{vap}=\sum_{i=0}^{z}dm_i$ that has entered the line equals the liquid mass $m_{liq}=\rho_{liq}V_{line}$ that can be contained in it, the transducer at the end of the line should start reading the correct pressure. The time instant at which this occurs can be determined by numerically solving the equation $m_{vap}=m_{liq}$ for the value of $z$. Time delay due to mass sink $\Delta \tau_{ms} = z \Delta t $ is proportional to vapor density, speed of sound and to minimum area size. 
Thus, it is expected to decrease for larger orifices, at higher pressure levels and lower Mach numbers due to the higher local density at the pressure tap.\\
Figure \ref{fig-Pline} shows that time delay predicted with the above procedure for \textit{Test B} is $\Delta \tau_{ms}= \SI{8.2}{s}$ and is in good agreement with experimental data having $\Delta \tau_{exp} \sim \SI{9.6}{s}$, confirming the mass sink mechanism occurrence. 
The underestimated prediction is attributable predominantly to the choked flow assumption. In reality, as the line fills and its pressure gets closer to the flow one, inflow into the pneumatic system will no longer be choked and the system will fill more slowly. \\
Considering \textit{Test A} on the total pressure tap, calculated time delay due to mass sink is negligible as pressure signals are always within measurement uncertainty of one another. This is due to the higher density at the total pressure tap and, most of all, to the much larger $A_{min}$. A roughly ten-fold increase in the minimum diameter corresponds to a two orders of magnitude decrease in time delay.\\ 
\indent If steady state measures were performed, it would be possible to operate with the line full of liquid. However, the TROVA operates in transient mode due to its batch nature so steady measurements are not possible. 
Thus, a line nitrogen flushing solution is put in place to overcome line condensation and avoid time delay due to mass sink effects. 


\section{Pneumatic System with Nitrogen Flushing}
\label{sec_PLflushN2}

\begin{figure}
	\begin{center}
		\includegraphics[width=1\textwidth,trim={0 0 0 0}, clip]{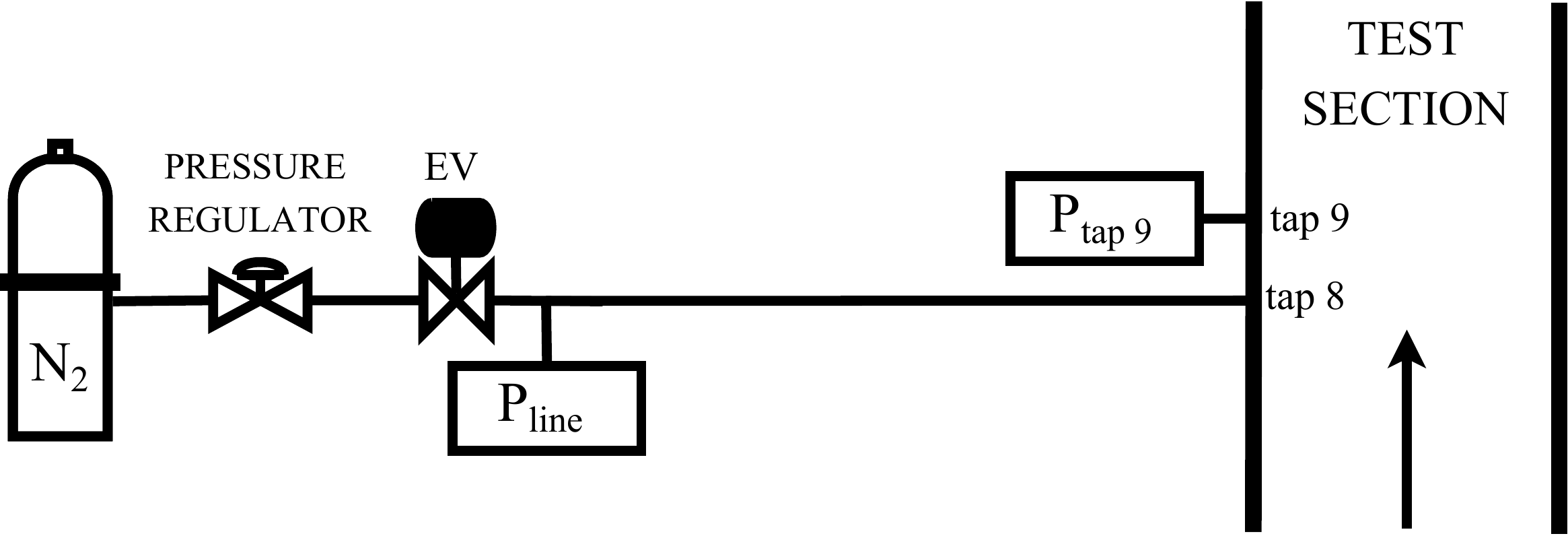}
		\caption{Single pneumatic line with nitrogen flushing.}
		\label{fig-N2flush}
	\end{center}
\end{figure}

\begin{figure}
	\begin{center}
		\includegraphics[width=0.7\textwidth,trim={0 0 0 0}, clip]{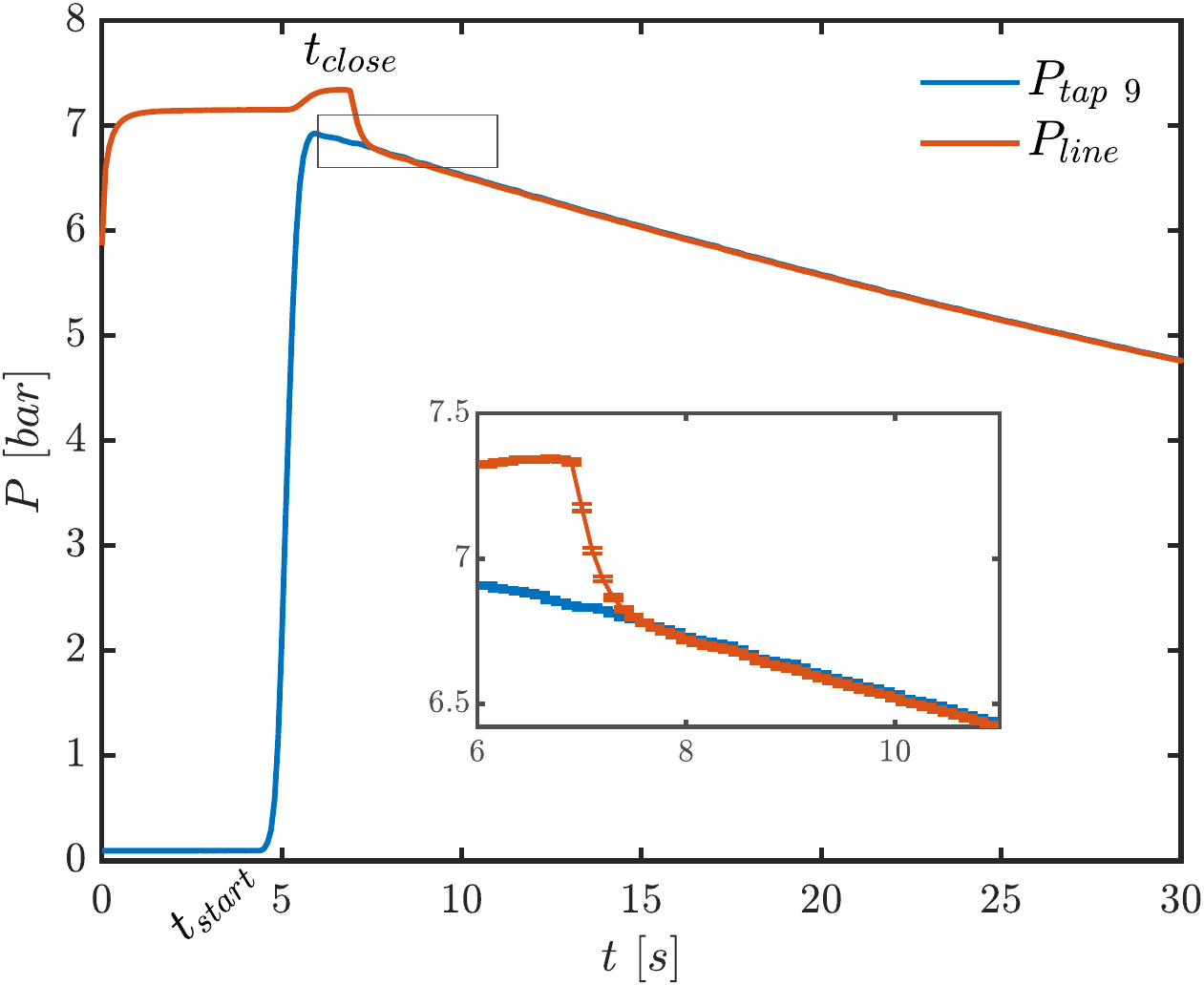}
		\caption{Pressure measures during \textit{Test C}. The first $\SI{30}{s}$ only are shown here for plot clarity.}
		\label{fig-PlineFlush}
	\end{center}
\end{figure}

\begin{figure}
	\begin{center}
		\includegraphics[width=1\textwidth]{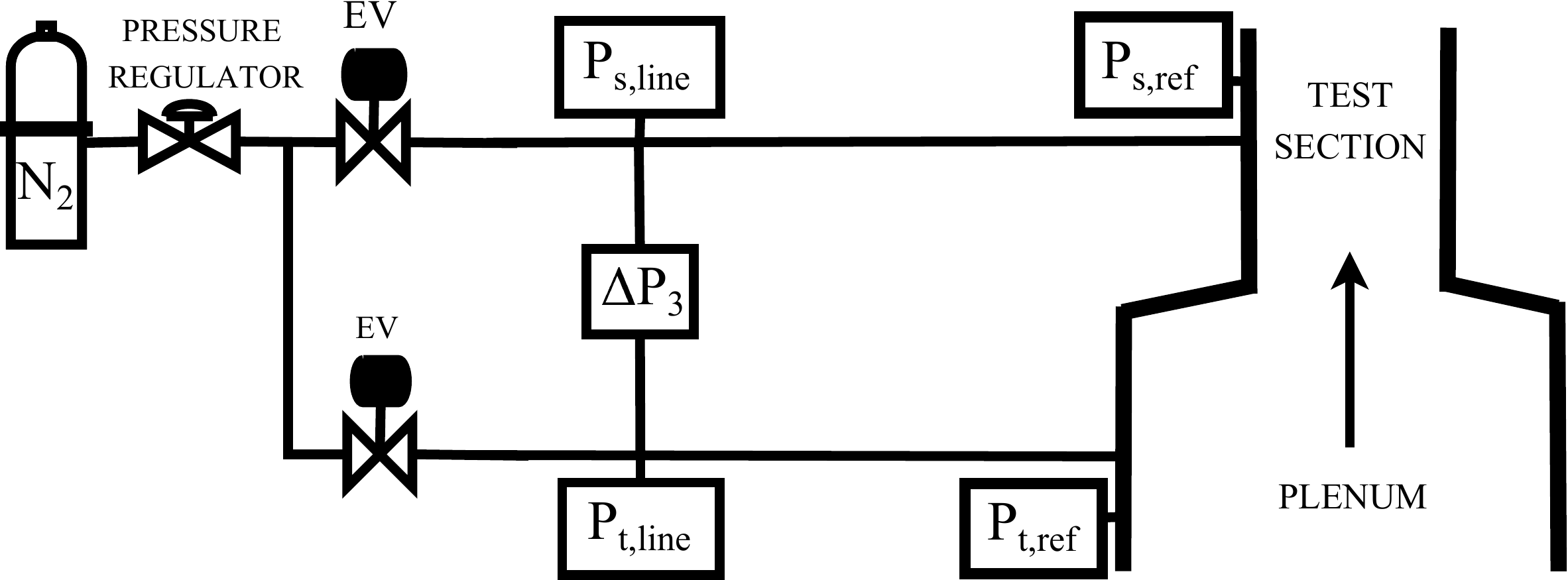}
		\caption{Differential pneumatic system with nitrogen flushing.}
		\label{fig-N2flushDiff}
	\end{center}
\end{figure}

\begin{figure}
	\begin{center}
		\includegraphics[width=0.7\textwidth]{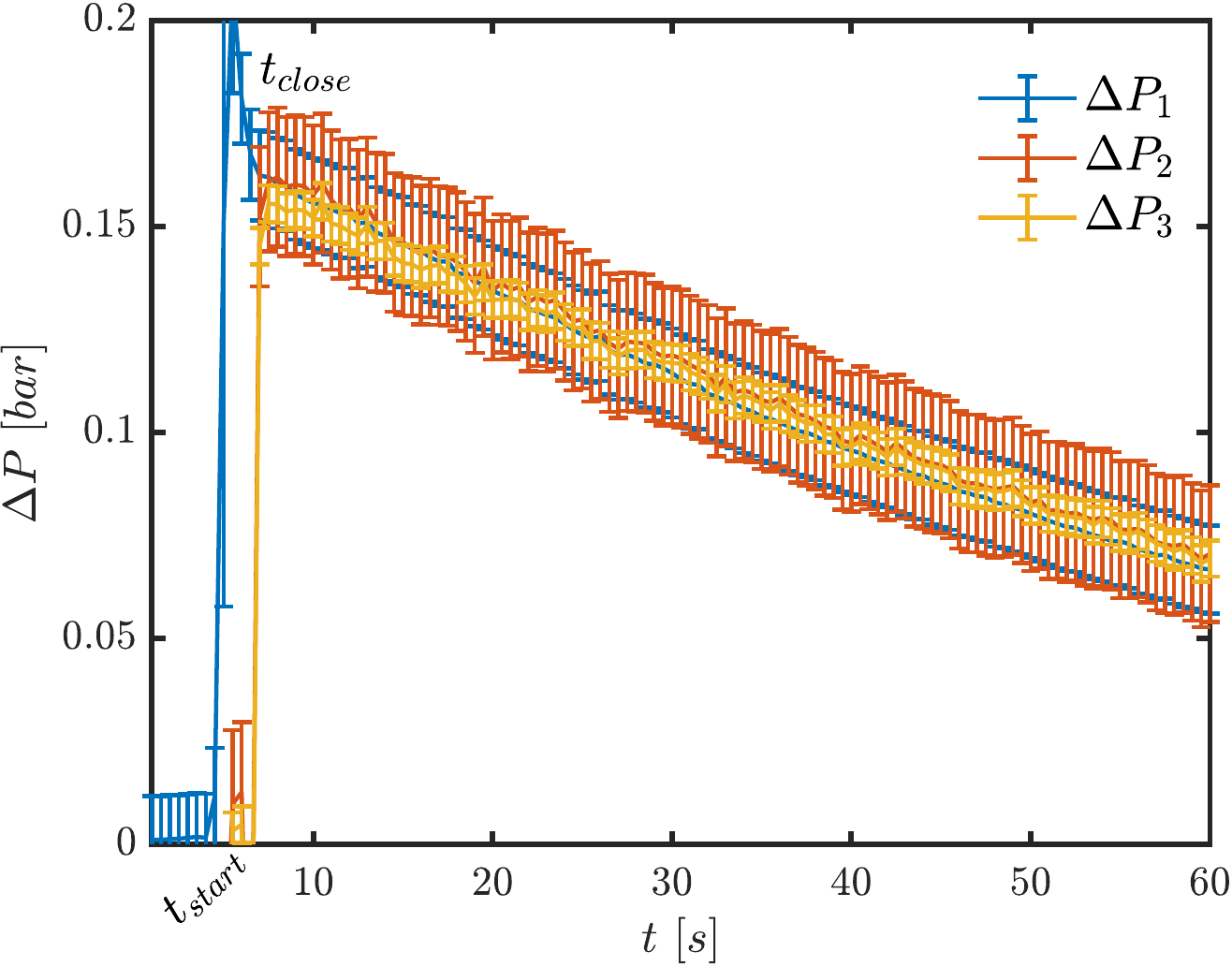}
		\caption{Kinetic head measurement during \textit{Test D}. }
		\label{fig-PlineFlushDiff}
	\end{center}
\end{figure}

The scheme of preliminary tests with one flushed line connected to a static wall tap is shown in Figure \ref{fig-N2flush}.
The line is directly connected to nitrogen storage tanks and pressure is regulated through a pressure reducer to just above the maximum expected one during the test. The line pressure transducer is mounted just downstream a solenoid electrovalve (\textit{EV}) at a distance of $ \sim \SI{1}{m}$ from the measurement point. Line length is here representative of the longest expected lines for future Pitot tubes testing. 
The valve is actuated by a \textit{Labview}\textsuperscript{\textregistered} program to open as the test is triggered and close right after the pressure peak is reached in the test section. This ensures that the line only contains nitrogen at all times during a test and no MM vapor enters it, so as to avoid condensation. As the test proceeds, nitrogen exits the line from the static tap into the test section as line pressure tends to equilibrium with the decreasing test section one.
Results of exemplary \textit{Test C} on nozzle $cMM02$ with initial total conditions $P_T=\SI{7}{bar}$ and $T_T=\SI{200}{\degreeCelsius}$ ($Z_T = 0.79$) are here reported. The line was mounted in the constant area region on tap $8$ and pressure signals were compared with a sensor mounted on the adjacent active tap $9$. \\ 
As Figure \ref{fig-PlineFlush} shows, pressure in the line decreases and is within error bars from flow pressure in well under $\SI{1}{s}$ after the solenoid valve is closed at $t_{close}$. Even lower time delay could be achieved by fine-tuning line nitrogen pressure and $t_{close}$ parameters. This result is an excellent improvement with respect to measurements involving condensation in the line, especially considering that the line in \textit{Test C} is about five times longer than in \textit{Test B}. Indeed, tests with condensation in the same \textit{Test C} configuration (not reported here for brevity) showed a time delay $\sim \SI{60}{s}$.\\
As previously explained, probes calibration requires differential pressure measurements to minimize uncertainty. Preliminary tests on a nitrogen flushed pneumatic line for differential pressure measures were carried out with the configuration shown in Figure \ref{fig-N2flushDiff}. Overall line length is about $\SI{1}{m}$ in view of future Pitot tubes testing. 
One line was installed at the total pressure tap in the plenum and the other one at tap $8$ in the constant area region of the nozzle to reproduce a similar setup to differential pressure measures between reference TROVA and Pitot tube total pressures. Differential pressure will also be measured between the Pitot tube static port and static pressure wall taps, but lines length will be significantly shorter.  
The objective of these preliminary tests was to compare the reference kinetic head measured with wall-mounted transducers $\Delta P_1=(P_t-P_s)_{ref}$ with the reading $\Delta P_3$ of the differential transducer. The plate-mounted reference pressure sensor is placed at tap $7$ in the constant area section.\\
In order to better identify possible issues with either one of the lines, absolute pressure transducers were also mounted on each line and the pressure difference  $\Delta P_2=(P_t-P_s)_{line}$ was calculated. 
Figure \ref{fig-PlineFlushDiff} reports results for \textit{Test D} with same initial total conditions and nozzle as \textit{Test C}, showing the excellent overlap between kinetic heads measured in the three different ways. 
The much smaller error bars associated with $\Delta P_3$ highlight the importance of using a differential configuration instead of calculating pressure difference from absolute transducers with the related propagated uncertainty.
It can be concluded that the nitrogen-flushed pneumatic system is able to correctly measure in differential mode aswell, given the complete agreement between the differential transducer and absolute plate-mounted ones. 
This aspect is of fundamental importance to achieve accurate pressure probe calibration with a low experimental uncertainty in the future.

\section{Complete Pneumatic System for Pitot Tube Testing in Non-Ideal Flows of Siloxane MM Vapor}
\label{sec-LinesPitotSub}
\label{sec-PitotMMsub}

\begin{figure}
	\begin{center}
		\includegraphics[width=1\textwidth]{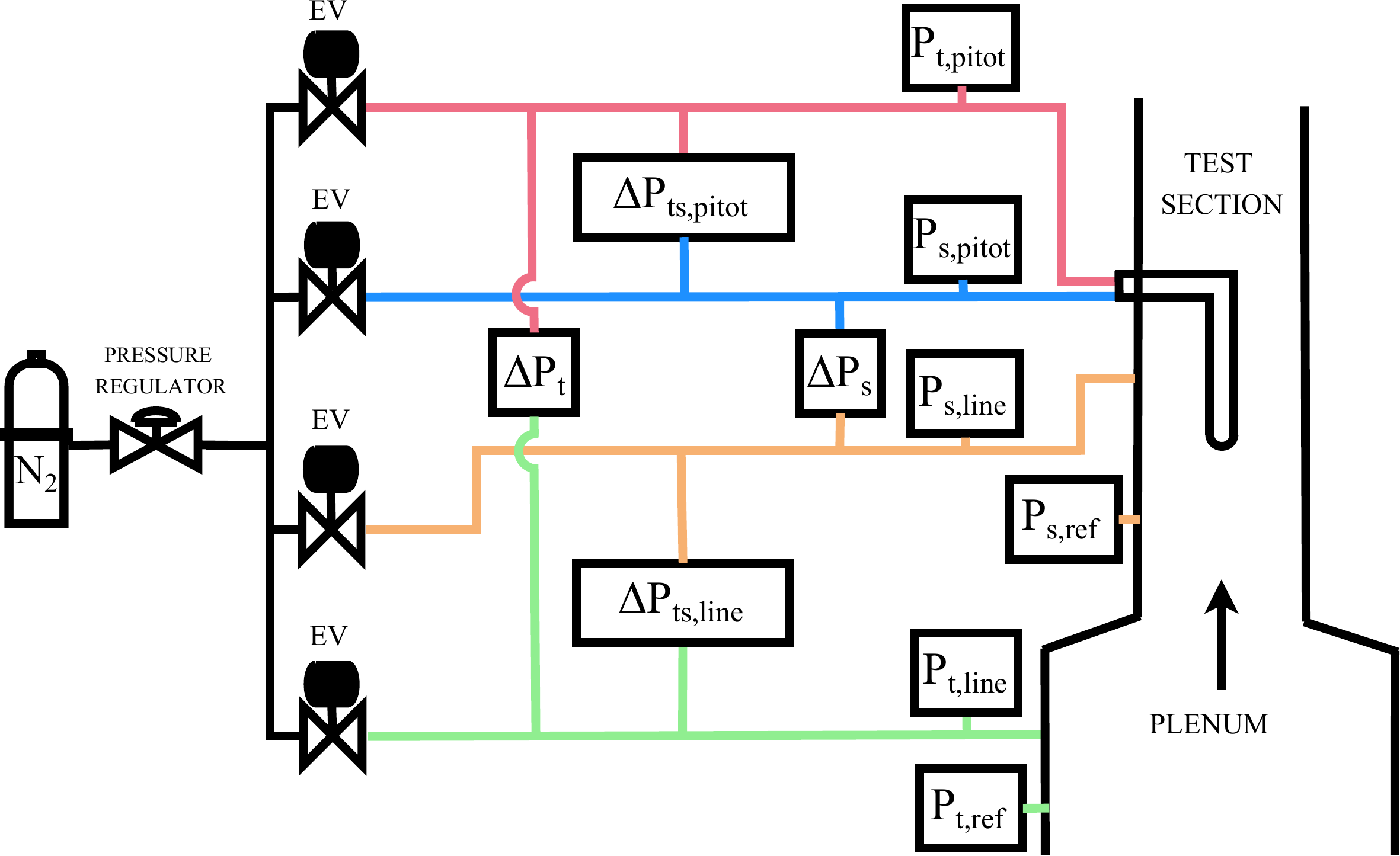}
		\caption{Flushed pneumatic lines scheme for Pitot tube testing in subsonic flows of Siloxane MM. Rectangular boxes represent pressure transducers.  }
		\label{fig-PlinePitotSub}
	\end{center}
\end{figure}

Building on the nitrogen flushed system developed and verified in the previous sections, the complete pneumatic lines configuration for testing with Pitot tubes is here presented. It is specifically devised using the \textit{Pitot} rear plate and includes several additional lines so that each quantity measured by the probe can be simultaneously acquired and compared against plant references, as shown in Figure \ref{fig-PlinePitotSub}.\\
The quantities of interest to be measured during \textit{TROVA} testing of Pitot tubes in subsonic flows of Siloxane MM are:
\begin{itemize}
	\item $P_{t,ref}$: reference total pressure in the TROVA plenum, measured with a wall-mounted absolute transducer;
	\item $P_{s,ref}$: reference static pressure in the constant cross section part of the subsonic choked nozzle, measured with a plate-mounted absolute transducer at tap number $2$ in Figure \ref{fig-convPitot};
	\item $P_{t,line}$: total pressure measured in the line exiting a pressure tap in the plenum;
	\item $P_{s,line}$: static pressure measured in the line exiting the wall tap in the constant cross-section region of the nozzle at tap number $4$ in Figure \ref{fig-convPitot}, in correspondance of the Pitot tube static ring;
	\item $P_{t,pitot}$: total pressure measured in the line connected to the total pressure port of the Pitot tube;
	\item $P_{s,pitot}$: static pressure measured in the line connected to the static ring of the Pitot tube;
	\item $\Delta P_{ts,line}$: reference kinetic head directly measured with a differential transducer between a total pressure tap in the plenum and static wall tap number $4$ in Figure \ref{fig-convPitot}.
	Given that this differential transducer is mounted on lines, its extremities are not exactly subject to $P_{t,ref}$ and $P_{s,ref}$, but actually to $P_{t,line}$ and $P_{s,line}$. This is why the subscript \textit{line} is used although this is taken as reference kinetic head of the plant;
	\item $\Delta P_{t}$: total pressure difference directly measured with a differential transducer between a pressure tap in the plenum and the Pitot tube and should thus correspond to $P_{t,line}- P_{t,pitot}$;
	\item $\Delta P_{s}$: static pressure difference directly measured with a differential transducer between static wall tap number $4$ and the Pitot tube. It should thus correspond to $P_{s,line}- P_{s,pitot}$;
	\item $\Delta P_{ts,pitot}$: kinetic head directly measured with a differential transducer by the Pitot tube. It   should thus correspond to $P_{t,pitot}- P_{s,pitot}$.
\end{itemize}


The last four quantities are acquired in differential mode to minimize
measurement uncertainty, as previously mentioned. 
Employed differential transducers are from the \textit{Schaevitz P2100} series, with full scale chosen in the range $0.7 - \SI{2.0}{bar}$ (uncertainty range $1.7 - 3.3~\permille$ of the full scale) according to the expected measured pressure difference.
Absolute transducers on each line measuring $P_{t,line}$, $P_{s,line}$, $P_{t,pitot}$ and $P_{s,pitot}$ are used as support to pinpoint any possible measurement issues.\\
Particular care was taken during system design and components positioning to minimize lines diameter, length, and fittings volume so as to decrease the overall response time as much as possible. \\

Commissioning of the complete pneumatic system including a probe was carried out with testing of an L-shaped Pitot tube with Siloxane MM vapor at nominal Mach numbers $M=0.2$ and $0.5$ using nozzles \textit{cMM02} and \textit{cMM05}, respectively.\\
One set of initial total conditions at mild/significant levels of non-ideality was considered for each nozzle as reported in Table \ref{tab-TROVAtestsSub}, together with the name of the chosen exemplary test for each considered condition. All results reported here are based on repeated testing with verified consistency and repeatability. 

\begin{table}
		\centering
		\begin{tabular}{ccccc}
			\hline
			\textbf{Name}  		 & \textbf{Nozzle}	& \textbf{$P_T$} & \textbf{$T_T $} & \textbf{$Z_T $}  \\ 
		                     	 &				  	& $[bar]$ 		 & $[^\circ C]$    & $[-]$ \\ \hline
			\textit{MM-0.2}		 &	\textit{cMM02}	& 7.36 & 202 &  0.78 \\ 
			\textit{MM-0.5}		 &	\textit{cMM05}  & 7.40 & 210 &  0.79 \\ \hline
		\end{tabular}%
		\caption{Initial total conditions of Pitot tube subsonic tests with siloxane MM. }
		\label{tab-TROVAtestsSub}
\end{table}

\subsection{$M=0.2$}
\begin{figure}
	\makebox[1\linewidth][c]{%
		\begin{subfigure}{0.55\textwidth}
			\centering	
			\includegraphics[width=1\textwidth]{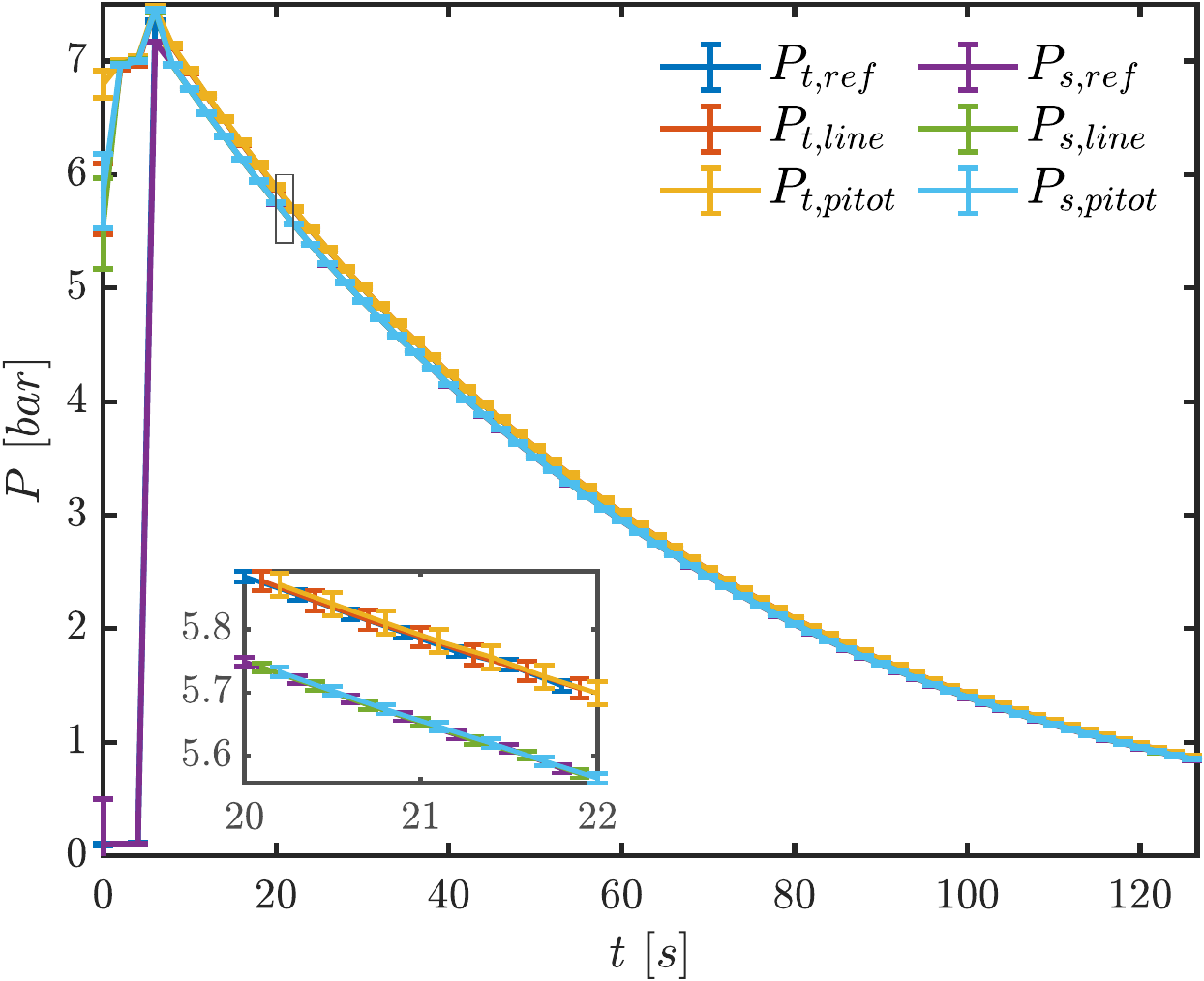}
			\caption{Absolute pressures. Quantities in the zoom box are at staggered instants for better visualization. }
			\label{fig-PitSubM02_1}
		\end{subfigure}
		\hspace{0.1pt}	
		
		\begin{subfigure}{0.55\textwidth}
			\centering	
			\includegraphics[width=1\textwidth]{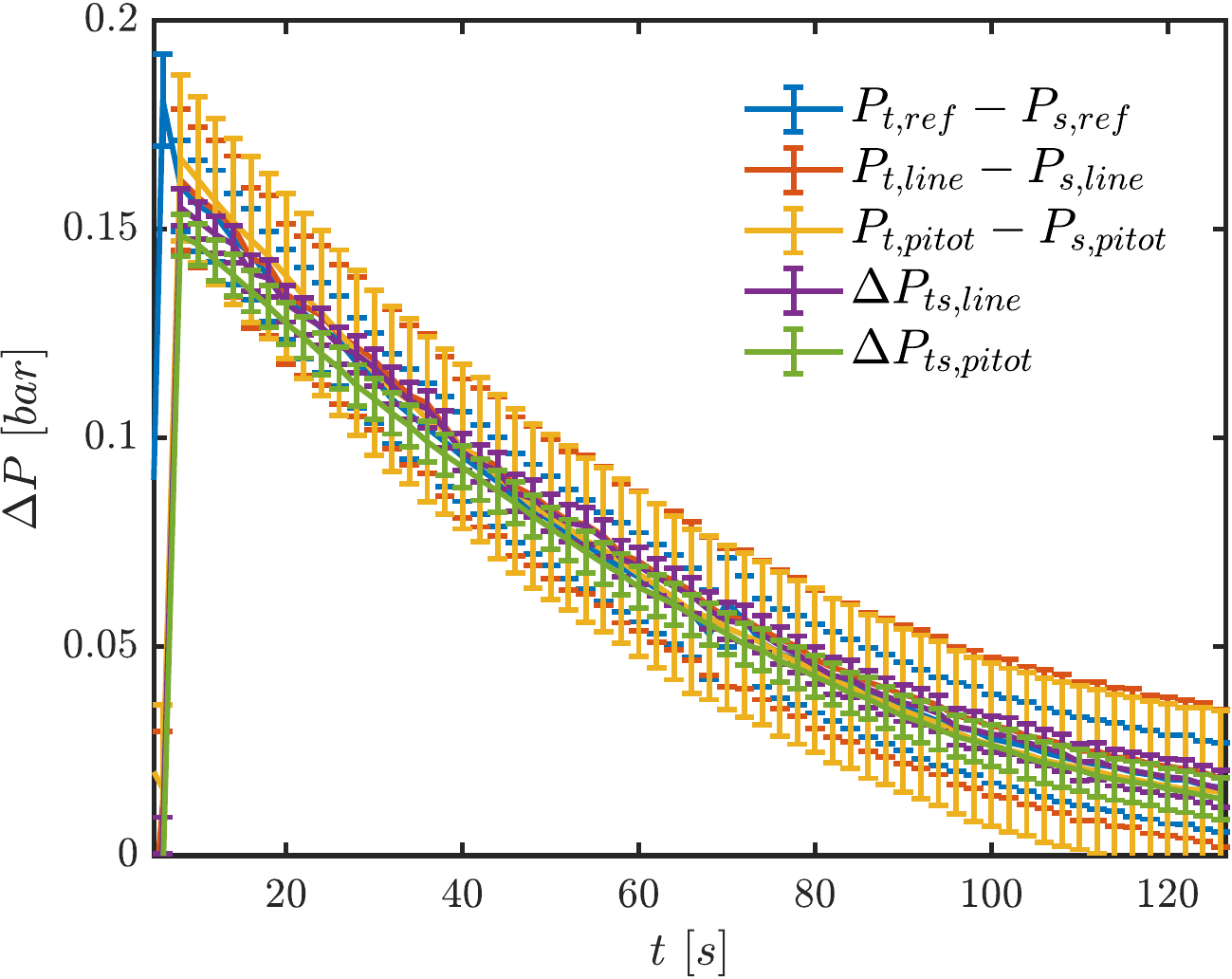}
			\caption{Kinetic head. \linebreak} 
			\label{fig-PitSubM02_2}
	\end{subfigure}	}
	
	\makebox[1\linewidth][c]{%
		\begin{subfigure}{0.55\textwidth}
			\centering	
			\includegraphics[width=1\textwidth]{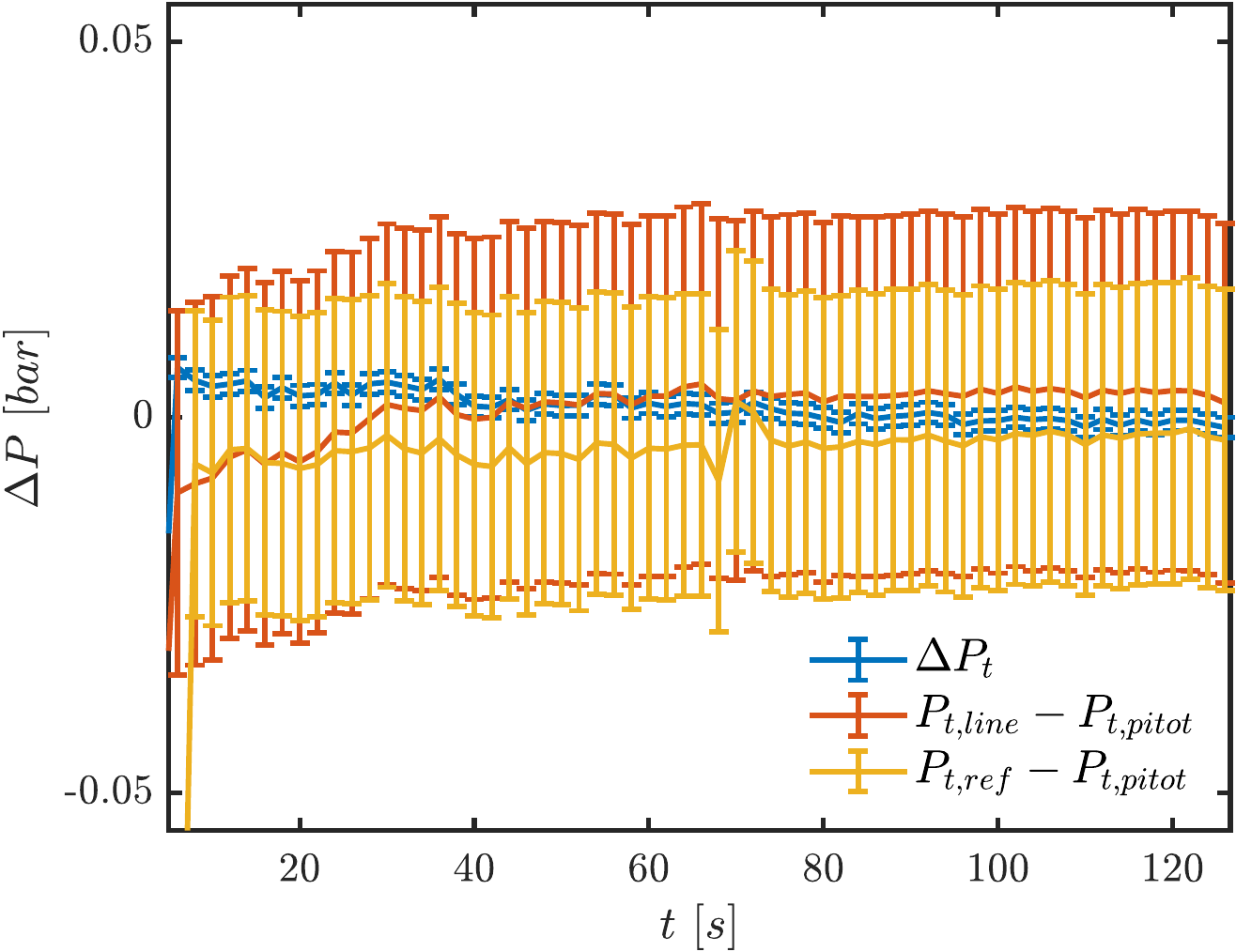}
			\caption{Total pressure difference.}
			\label{fig-PitSubM02_3}
		\end{subfigure}
		\hspace{0.1pt}	
		
		\begin{subfigure}{0.55\textwidth}
			\centering	
			\includegraphics[width=1\textwidth]{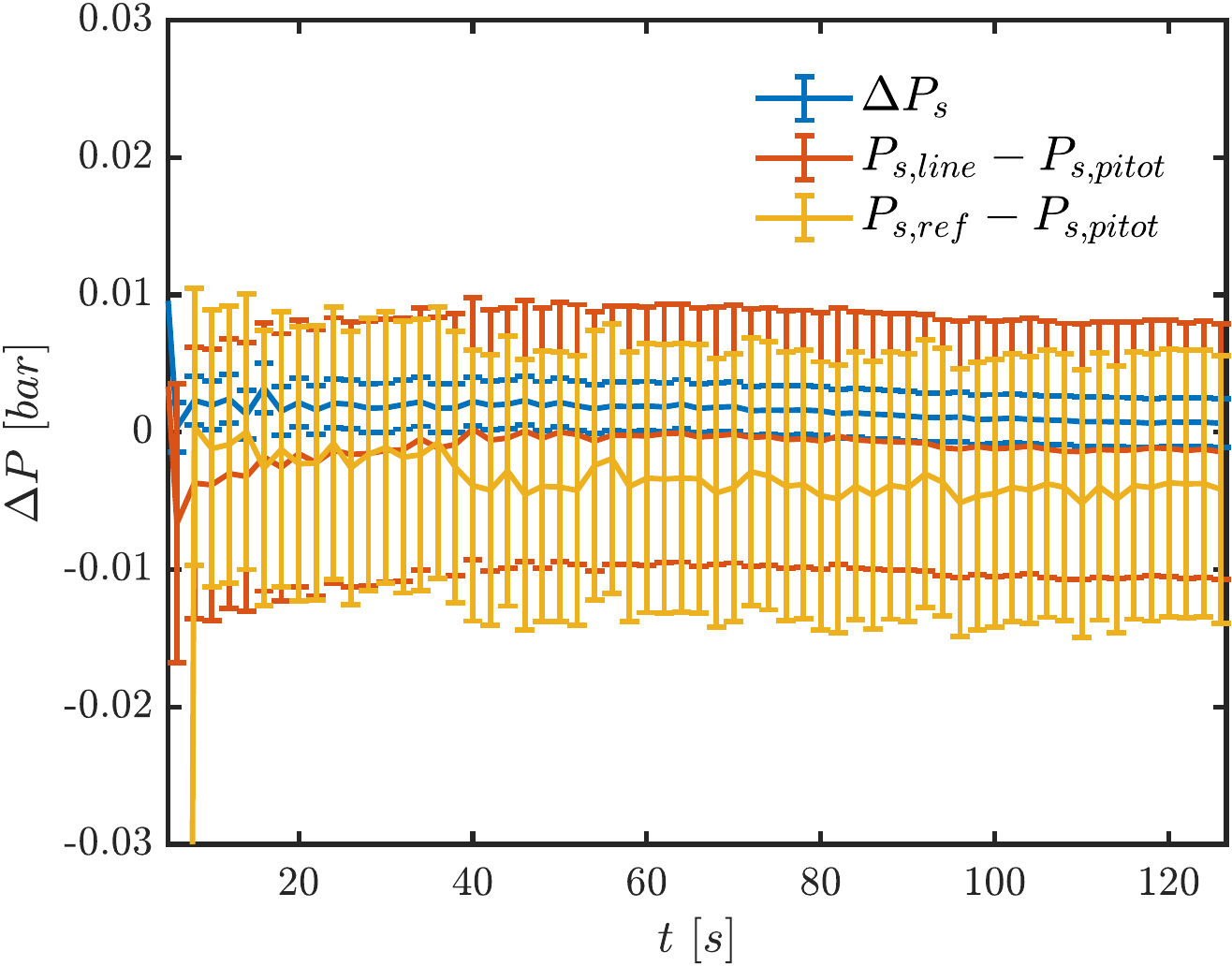}
			\caption{Static pressure difference.}
			\label{fig-PitSubM02_4}
	\end{subfigure}	}
	
	\caption{Results from test \textit{MM-0.2} with nozzle \textit{cMM02}. Pressure differences directly measured with differential transducers are also compared to the ones calculated from absolute transducers. } 
	\label{fig-PitSubM02}
\end{figure}

Results from test \textit{MM-0.2} presented in Figure \ref{fig-PitSubM02} allow to evaluate the system performance at subsonic conditions with a Mach number $M\simeq 0.2$.\\
All acquired absolute pressures are plotted as a function of time in Figure \ref{fig-PitSubM02_1}. 
The common nitrogen line upstream of the electro valves does not affect measurements. Any pressure equalization (from flushing pressure to the one at the corresponding measurement point in the test section for total, static and differential quantities) occurs within the sharp emptying transient of the lines, when electro valves are closed just after peak pressure is reached.\\
The reference total pressure measured with a wall-mounted transducer is superposed to $P_{t,line}$ and $P_{t,pitot}$. 
The total pressure difference $\Delta P_t$, directly measured with a differential transducer, is $\sim \SI{7}{mbar}$ at test beginning and decreases down to  $\SI{1}{mbar}$ at the end. This difference is below $3\%$ with respect to the flow kinetic head, confirming the satisfactory agreement.
This can be seen in Figure \ref{fig-PitSubM02_3}, where $\Delta P_t$ is reported together with pressure difference computed from absolute transducers. All three quantities agree, although the calculated pressure differences have a very large error-bar due to the large uncertainty propagated from absolute measures. This again highlights the importance of using differential transducers instead of absolute ones in the present case. \\
Quite analogously, the three measured static pressures $P_{s,ref},P_{s,line}$ and $P_{s,pitot}$ in Figure \ref{fig-PitSubM02_1} are also superposed. 
Consistently, the static pressure difference $\Delta P_s$ in Figure \ref{fig-PitSubM02_4} is only $\SI{3}{mbar}$ at test start (less than $1.5\%$ with respect to the kinetic head) and decreases to less than $\SI{1}{mbar}$ at the end, and its trend agrees very well with the difference between absolute transducers. \\
Given the good measurement performance in both total and static quantities, the kinetic head reported in Figure \ref{fig-PitSubM02_2} shows a perfect overlap between readings from all differential and absolute transducers. \\
Overall, results show the adequate performance of the complete pneumatic system in non-ideal flows of siloxane MM at $M\simeq0.2$ for total, static and kinetic head measurements. 

\subsection{$M=0.5$}

\begin{figure}
	\makebox[1\linewidth][c]{%
		\begin{subfigure}{0.55\textwidth}
			\centering	
			\includegraphics[width=1\textwidth]{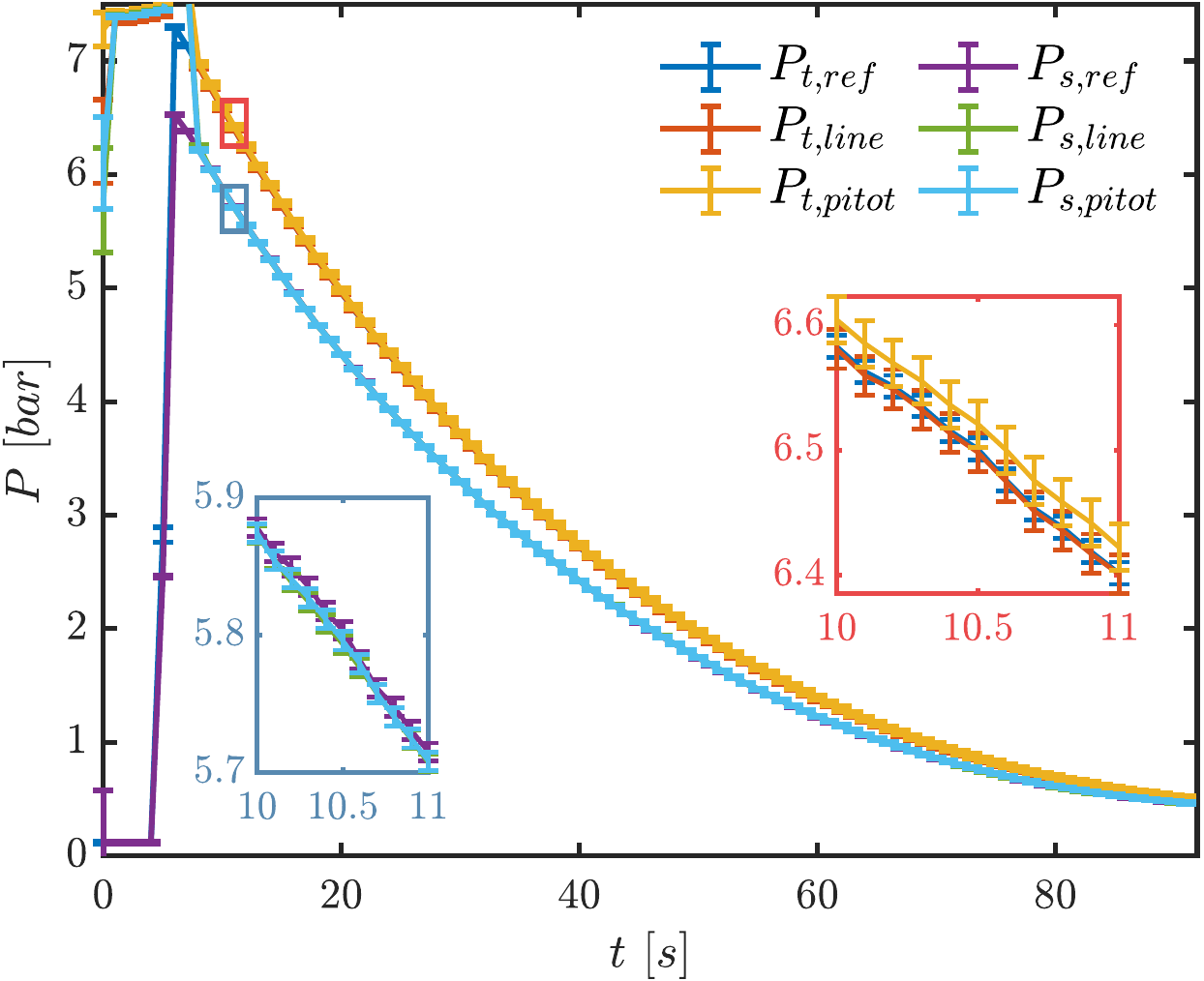}
			\caption{Absolute pressures.}
			\label{fig-PitSubM05_1}
		\end{subfigure}
		\hspace{0.1pt}	
		
		\begin{subfigure}{0.55\textwidth}
			\centering	
			\includegraphics[width=1\textwidth]{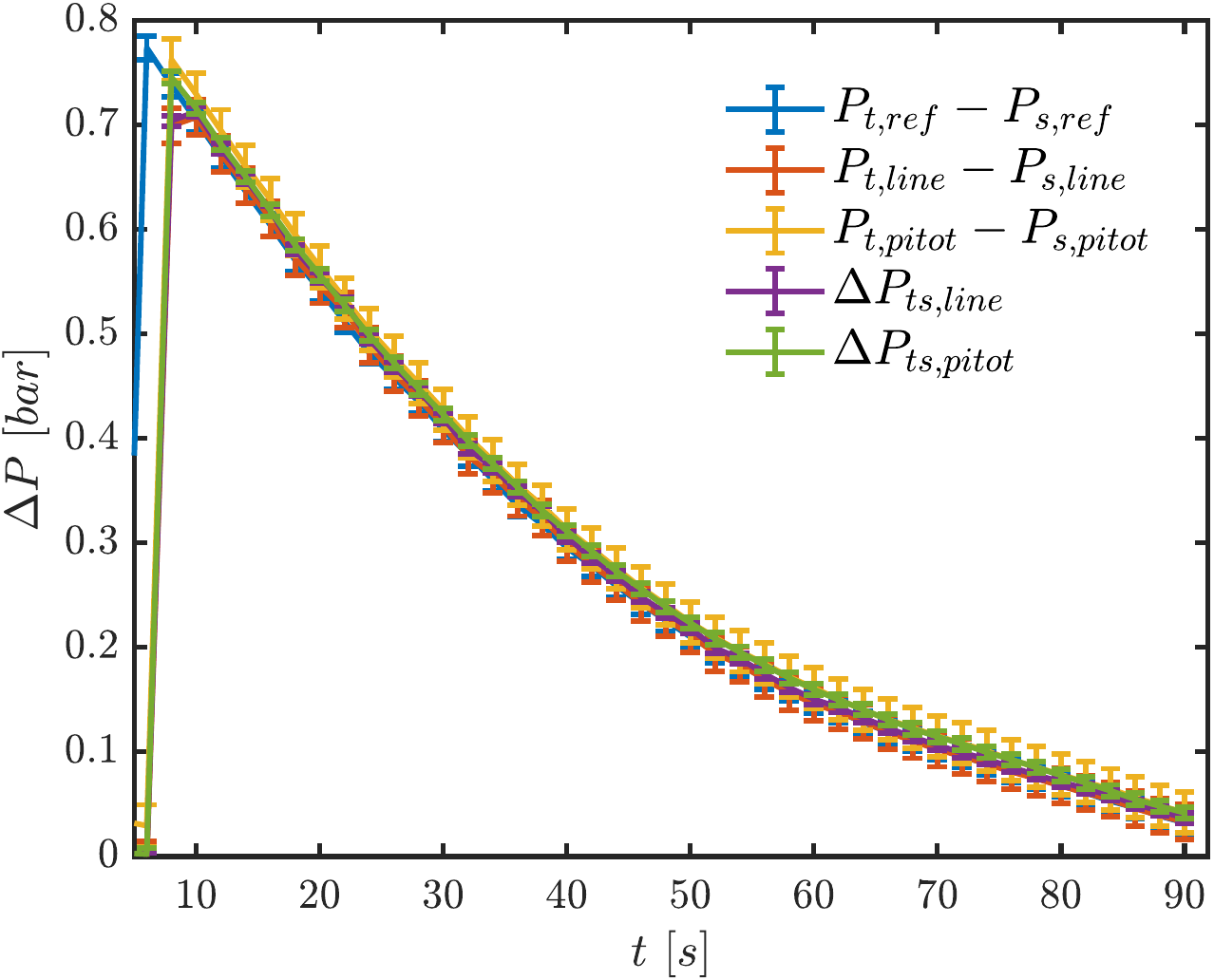}
			\caption{Kinetic head.}
			\label{fig-PitSubM05_2}
	\end{subfigure}	}
	
	\makebox[1\linewidth][c]{%
		\begin{subfigure}{0.55\textwidth}
			\centering	
			\includegraphics[width=1\textwidth]{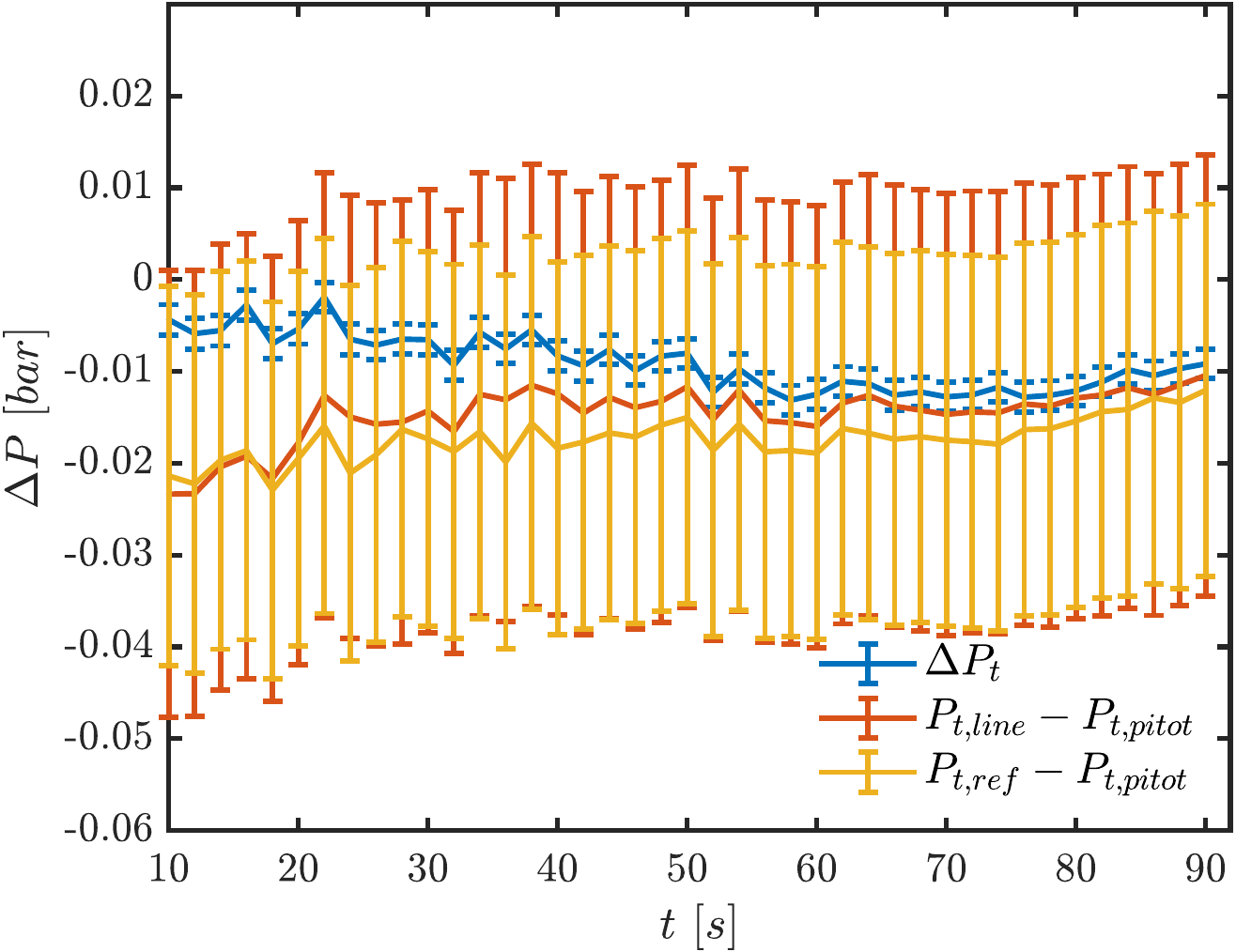}
			\caption{Total pressure difference.}
			\label{fig-PitSubM05_3}
		\end{subfigure}
		\hspace{0.1pt}	
		
		\begin{subfigure}{0.55\textwidth}
			\centering	
			\includegraphics[width=1\textwidth]{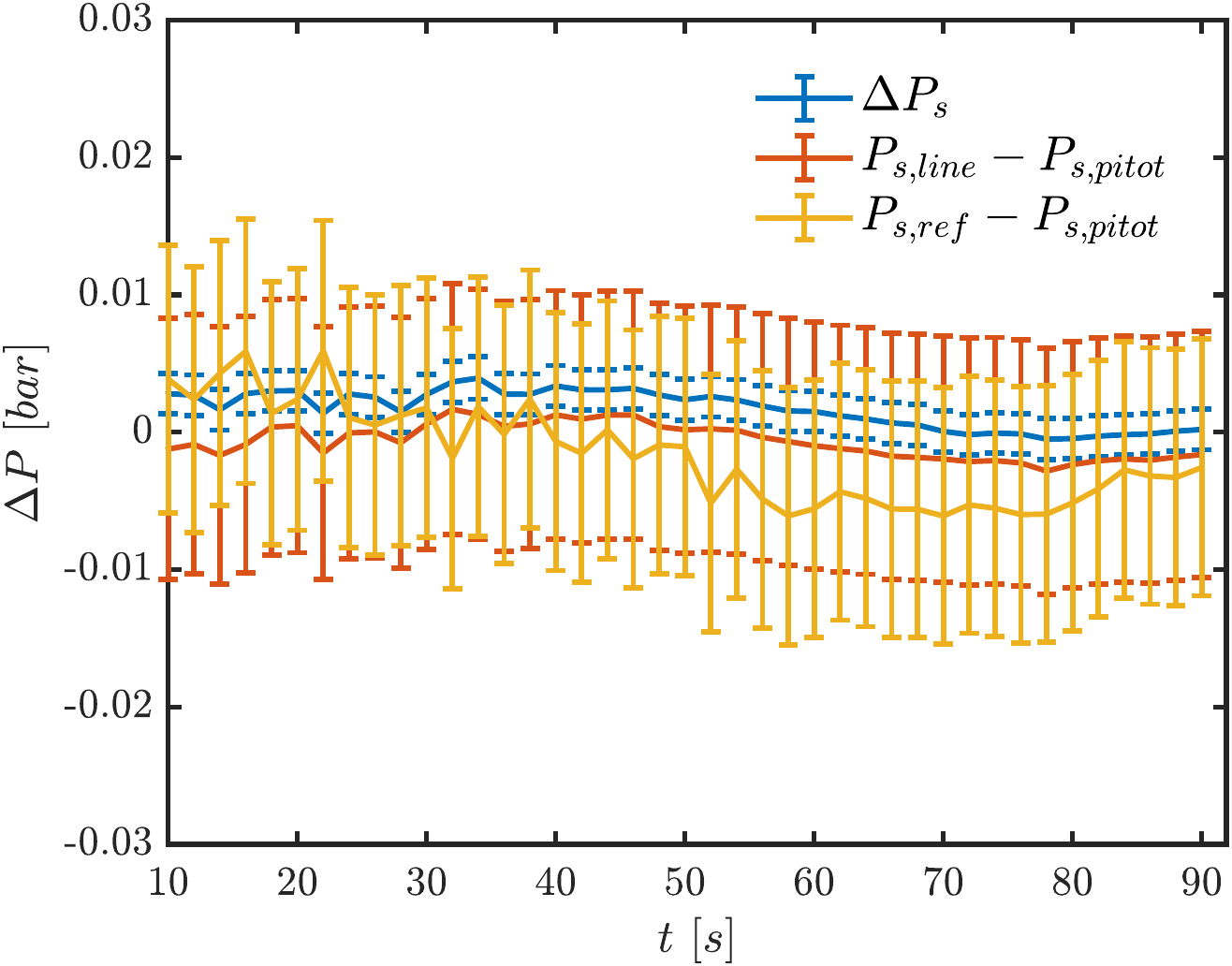}
			\caption{Static pressure difference.}
			\label{fig-PitSubM05_4}
	\end{subfigure}	}
		
	\caption{Results from test \textit{MM-0.5} with nozzle \textit{cMM05}. Pressure differences directly measured with differential transducers are also compared to the ones calculated from absolute transducers. } 
	\label{fig-PitSubM05}
\end{figure}

Figure \ref{fig-PitSubM05} reports results from test \textit{MM-0.5} with the complete pneumatic system operating in non-ideal flows at $M\simeq 0.5$.\\
Analogously to testing at $M\simeq 0.2$, the common nitrogen line upstream of the electro valves does not affect measurements. Pressure transient after their closure occurs again over a very short time span, which does not significantly differ between total and static lines even though, given the higher Mach number, static pressure is now significantly lower than flushing pressure.\\
However, differently with respect to testing at $M\simeq 0.2$, the Pitot tube total pressure $P_{t,pitot}$ is not well superposed with respect to reference and line total pressures after the end of the emptying transient linked to electro valves shutoff. This can be seen in the red zoom box in Figure \ref{fig-PitSubM05_1} where, although $P_{t,pitot}$ is still within errobars of the other two, it is overall markedly higher. 
This is reflected in a negative total pressure difference $\Delta P_t$ in Figure \ref{fig-PitSubM05_3} with an average value of $\sim \SI{-10}{mbar}$ that, towards the end of the test, corresponds to even $\sim 10\%$ of the kinetic head.
The negative measure by the differential transducer is confirmed by the difference between absolute pressures, albeit with much larger uncertainties. \\
Considering static pressures in Figure \ref{fig-PitSubM05_1}, $P_{s,ref}$, $P_{s,line}$ and $P_{s,pitot}$ are all well within errorbars of one another, consistently with a $\Delta P_s$ value in Figure \ref{fig-PitSubM05_4} always well below $\SI{3}{mbar}$, corresponding to a percentage difference with respect to the kinetic head lower than $1\%$.\\
To conclude, it is evident that the pneumatic system is adequate for static pressure measures, but an issue on the Pitot tube total pressure line is present, as further investigated next. 

\subsection{Measurement Delay}
\label{sec-PitotAPdelay}

\begin{figure}
	\begin{center}
		\includegraphics[width=0.7\textwidth]{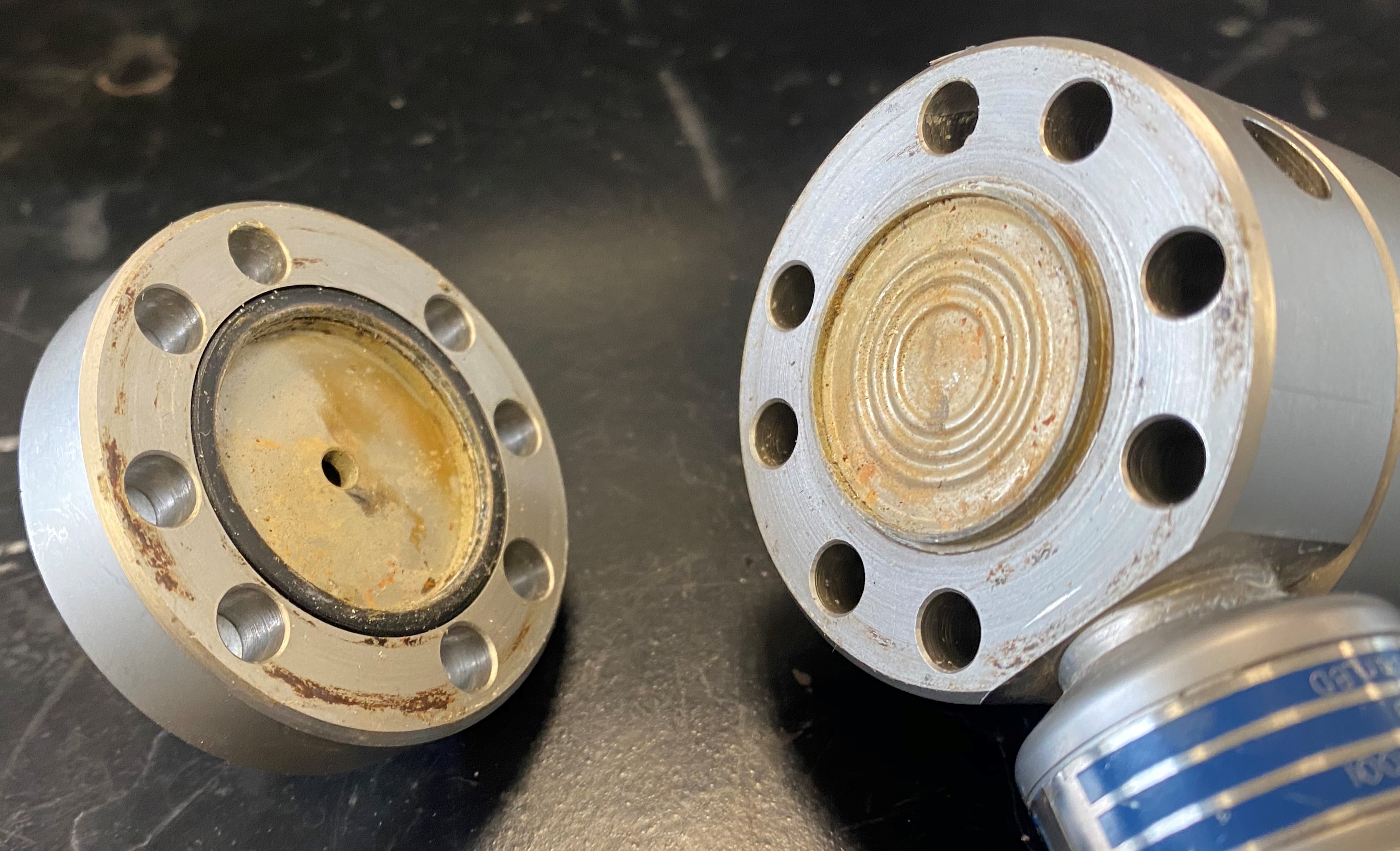}
		\caption{Internal membrane chamber of a \textit{Schaevitz} differential pressure transducer.}
		\label{fig-Schaevitz_interno}
	\end{center}
\end{figure}

Pitot tube testing with nozzle \textit{cMM05} highlighted that its total pressure $P_{t,pitot}$ is larger than the total reference and line ones.
The only physically possible explanation for this, which is also consistent with the emptying dynamics in which the pneumatic line operates, is of a measurement delay. 
Line pressure is unable to readily adapt to the decreasing pressure in the test section, meaning that at any given time instant, the pressure in the total pressure line of the Pitot tube is larger than the reference one, leading to errors in both absolute and differential measures. \\
The extent of the delay was found to increase when nozzles with larger throat area were employed (\textit{cMM05} with respect to \textit{cMM02}, in which case the delay was not noticeable) so when the emptying dynamics of the $HPV$ was faster.\\
This delay was somewhat unexpected, given that no such issues were evident during preliminary lines testing without the probe, previously described. 
Given that the Pitot tube was indeed absent during such tests, the delay was initially attributed to a possible contraction in its total pressure inner channel. However, 
probe inspection indicated that the line was actually clear. The reason for the measurement delay is instead to be found in a different overall lines volume between the pneumatic systems used for testing without and with the probe. 
This was because the former preliminary campaign was carried out with nozzle \textit{cMM02}, involving a slow emptying \textit{HPV} dynamics, and with only one differential transducer on each line. The complete system for probes testing instead features two differential sensors on each line, one for the kinetic head and the other measuring the difference between Pitot tube and line static or total pressures. As evident in Figure \ref{fig-Schaevitz_interno}, the employed differential transducers are characterized by a relatively large internal volume in the membrane chamber leading to a hidden but significant increase in the overall lines volume.
This dramatically increases the pneumatic line nitrogen discharge time during tests and is responsible for the larger total pressure measured by the Pitot tube. \\
Pneumatic system dynamic testing, described in the following section, allowed to confirm the source of the unwanted measurement delay.\\ 
It should be pointed out that delay is instead not present in the measure of $P_{t,line}$, even though the same number and type of transducers are mounted on the related line too.
This is due to the larger minimum diameter ($\SI{2}{mm}$), which ensures that the line can empty much faster than the Pitot tube total pressure one having a tip discharge orifice of $\SI{0.6}{mm}$ in diameter. \\

\section{Pneumatic Lines Dynamic Testing and Reduced Delay}
\label{sec-pnemDynTest}
\label{sec-RedDelay}

\begin{figure}
	\makebox[1\linewidth][c]{%
		\begin{subfigure}{1\textwidth}
			\centering	
			\includegraphics[width=1\textwidth]{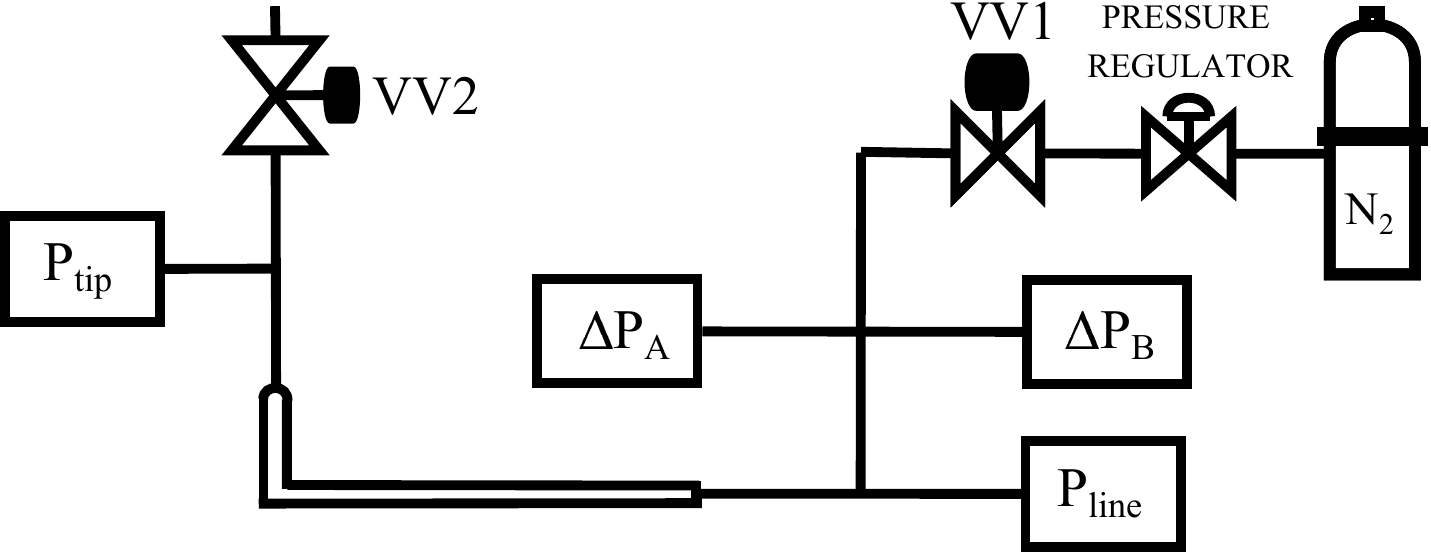}
		\end{subfigure}
	}
	\caption{Pneumatic lines scheme for dynamic testing of the probe total pressure line. Differential transducers $\Delta P_A$ and $\Delta P_B$ can be \textit{Schaevitz} or \textit{Kulite} sensors.}
	\label{fig-tardinFotoSchema}
	\label{fig-tardinScheme}
\end{figure}

\begin{figure}
	\makebox[1\linewidth][c]{%
		\begin{subfigure}{0.55\textwidth}
			\centering	
			\includegraphics[width=1\textwidth]{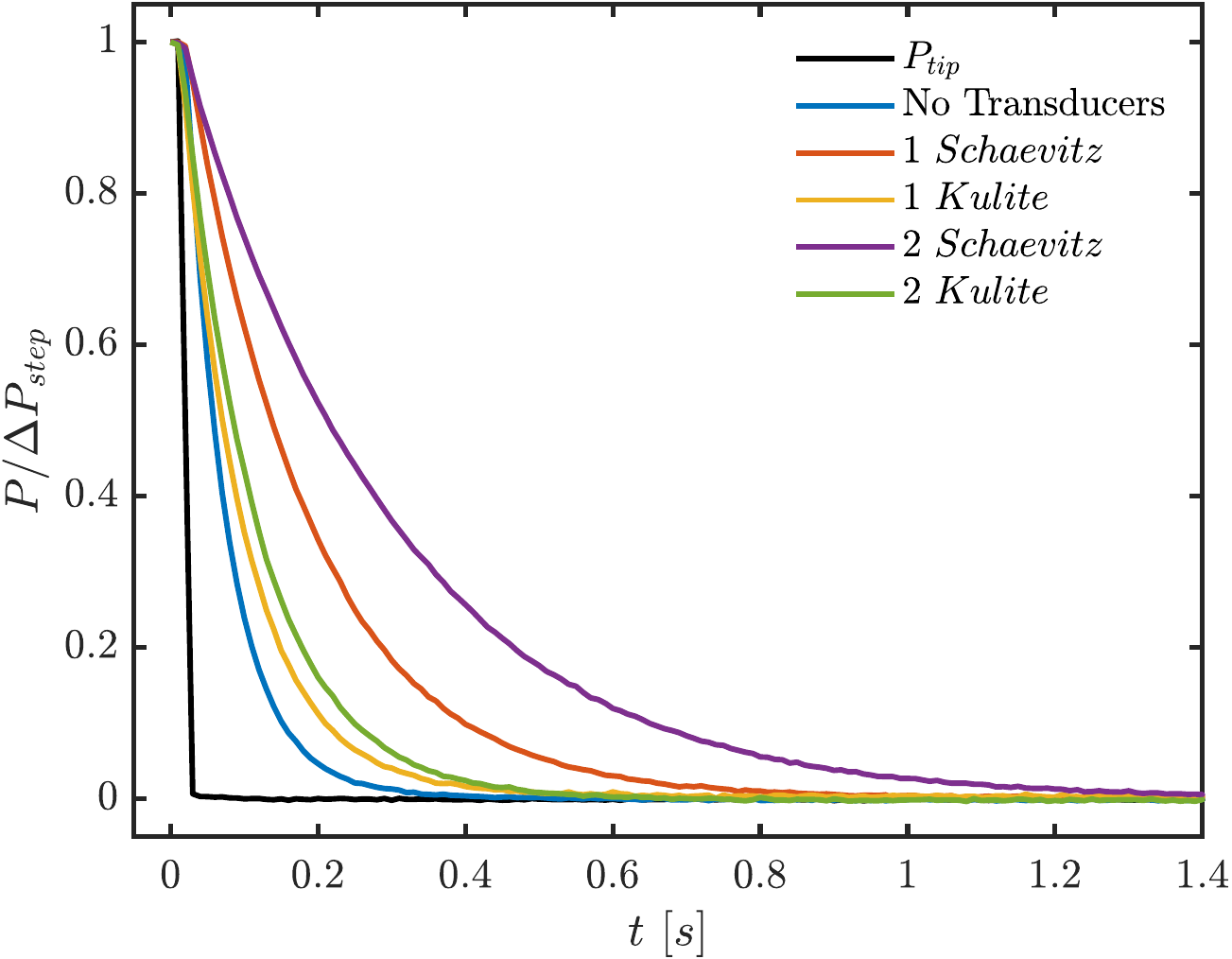}
			\caption{Step pressure response. }
			\label{fig-step}
		\end{subfigure}
		\hspace{0.1pt}	
		
		\begin{subfigure}{0.55\textwidth}
			\centering	
			\includegraphics[width=1\textwidth]{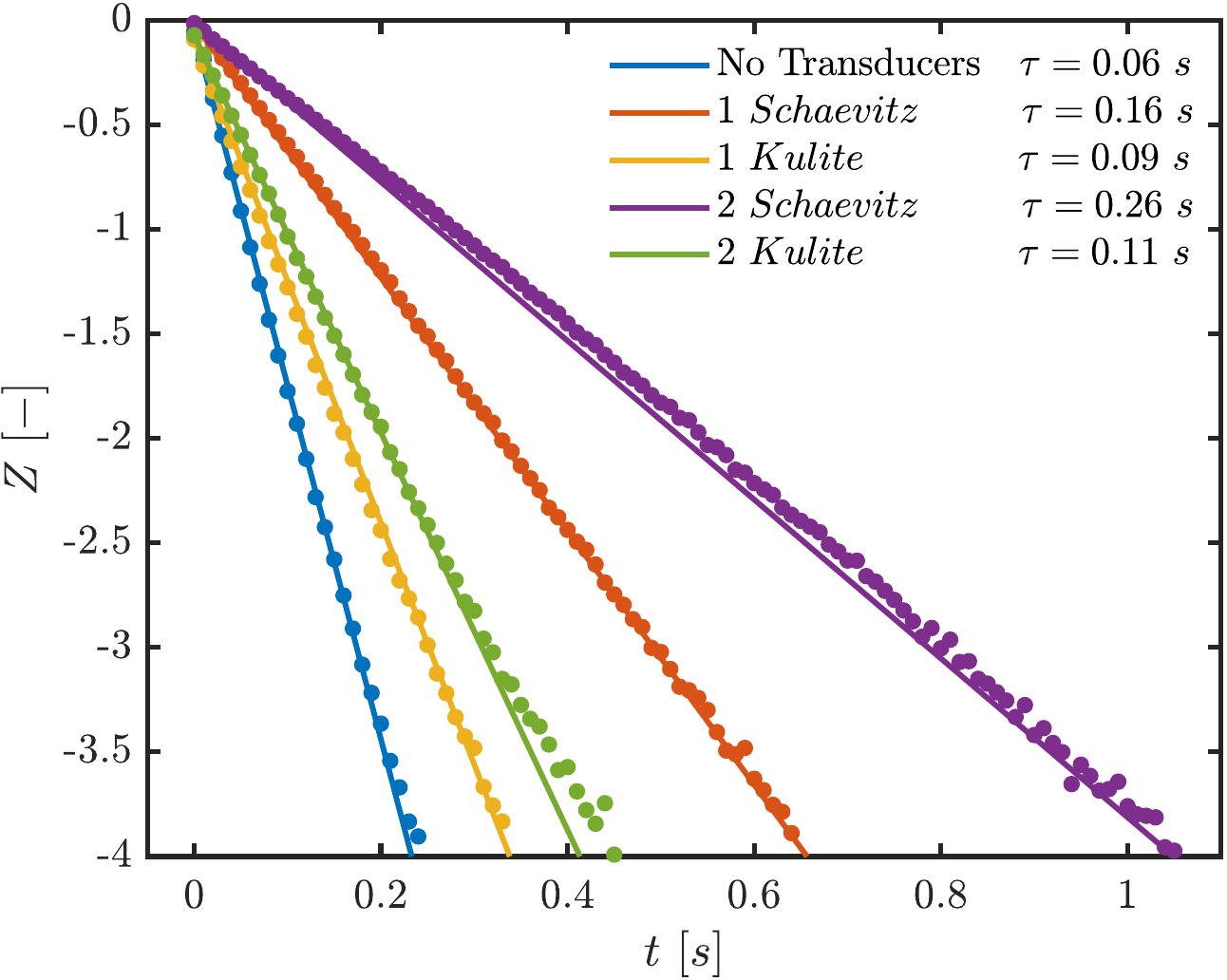}
			\caption{Quantity $Z$ as a function of time and data fitting by straight lines with slope $-1/\tau$. }
			\label{fig-tau}
		\end{subfigure}	
	}
	\caption{Dynamic testing of several total pressure line configurations, starting at $t=0~s$, when electrovalve \textit{VV2} is opened. } 
\end{figure}

Just before a typical \textit{TROVA} test start, pneumatic lines are loaded with nitrogen. It is then discharged into the test section as the lines empty as the test proceeds, due to the transient nature of the plant. Therefore, the aim of the dynamic testing procedure is to assess the Pitot tube total pressure line response as it empties. \\
To this end, the scheme shown in  Figure \ref{fig-tardinScheme} is employed. The probe total pressure pneumatic system, complete with electrovalve \textit{VV1} and absolute and differential transducers, is the same as during actual tests so as to account for the real lines volume. The other end of differential transducers $\Delta P_A$ and $\Delta P_B$ is left open to ambient conditions. 
An absolute reference transducer and an electrovalve \textit{VV2} are added at the probe tip. \\
The dynamic testing procedure is as follows. The total pressure line is pressurized by opening electrovalve \textit{VV1} while \textit{VV2} is closed. \textit{VV1} is then closed after a desired constant pressure is reached. In this case, line pressurization up to $\sim\SI{250}{mbar_r}$ was considered sufficient, since line discharge during tests occurs over very small pressure differences with the test section (except at the very first test instants just after electrovalves closure).
\textit{VV2} is then opened, imposing a negative input to the line as it discharges into the atmosphere from the probe tip, analogously to what happens during test time. Pressure read by the absolute transducer at the end of the line $P_{line}$ is compared to the reference absolute transducer at the probe tip $P_{tip}$, as shown in Figure \ref{fig-step} in order to determine the line response. \\
It must be pointed out that electrovalve \textit{VV2} is not a fast-opening valve. However, its opening time is orders of magnitude lower than the characteristic time of pressure decrease due to \textit{HPV} emptying during a test. It was therefore considered sufficiently fast to assess the dynamic response of the system in the present case.


To optimize the Pitot tube total pressure line response time, various line configurations were also tested, featuring two, only one or no mounted differential transducers. 
Given that the excessive chamber volume of the employed \textit{Schaevitz} transducers is the reason for measurement delay, 
\textit{Kulite} differential sensors of \textit{XTL-3-375 (M)} series were identified as candidate substitutes given their significantly smaller size. Therefore, the lines dynamic response was also tested in configurations including the latter.\\
Figure \ref{fig-step} reports the response for all tested line configurations, as indicated in the legend. These include cases with no mounted sensors as well as one or two transducers of each type. 
A $\sim\SI{250}{mbar_r}$ step was applied in all cases, but the plotted pressure trend was normalized with respect to the exact imposed one for comparison purposes.  
The value of $P_{line}$ is plotted for all configurations, as well as the tip pressure $P_{tip}$ for reference. \\
Results quite self-evidently provide the explanation for the measurement delay issues identified on the Pitot tube total pressure line.
The use of two \textit{Schaevitz} differential transducers significantly slows down the system response, given that the line pressure does not reach the tip one before $\SI{1.4}{s}$ after the step is applied. This is significantly later with respect to the configuration with one transducer only, which indeed was the one used during preliminary pneumatic line testing prior to probe insertion in the test section. \\
\textit{Kulite} transducers provide instead much less measurement delay and are therefore preferred for pressure probes testing here, instead of the initial \textit{Schaevitz} ones. It must be pointed out that the two types of sensor feature a comparable accuracy and the producer choice was only determined by the readiness of available models and their compatibility with siloxane fluids.

\begin{figure}
	\makebox[1\linewidth][c]{%
		\begin{subfigure}{0.55\textwidth}
			\centering	
			\includegraphics[width=1\textwidth]{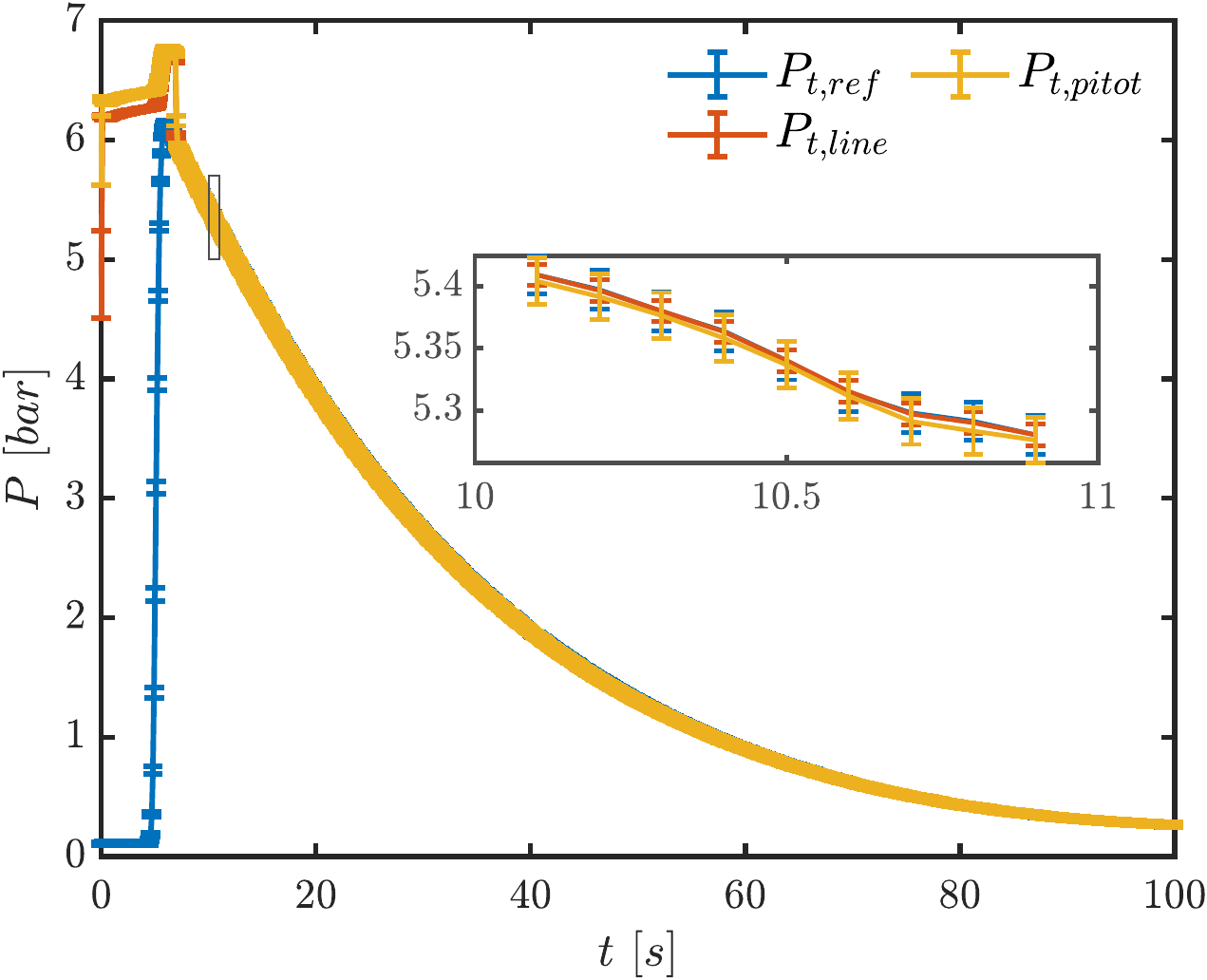}
			\caption{Absolute total pressures.}
			\label{fig-PitSubM05_MM206_1}
		\end{subfigure}
		\hspace{0.1pt}			
		\begin{subfigure}{0.55\textwidth}
			\centering	
			\includegraphics[width=1\textwidth]{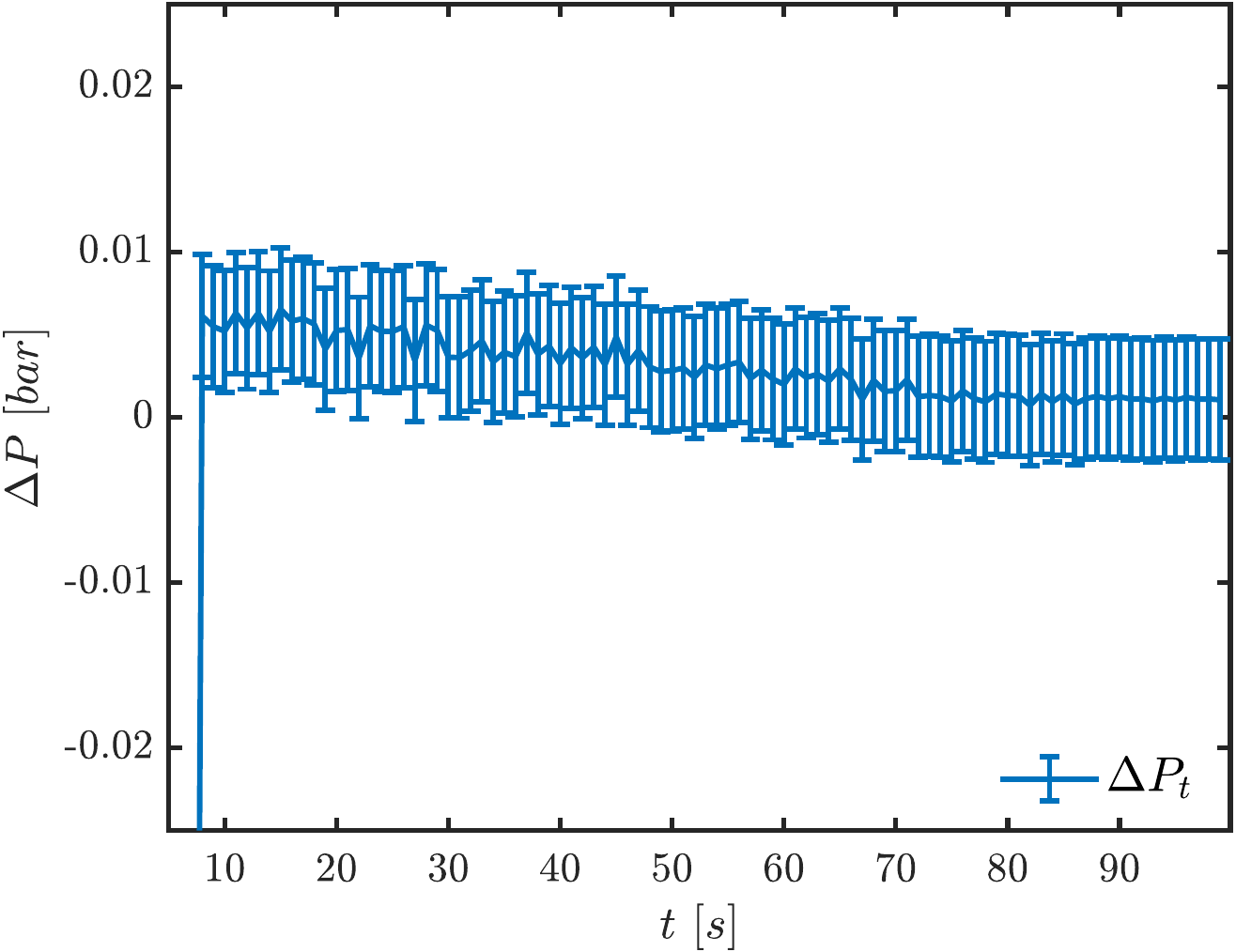}
			\caption{Total pressure difference.}
			\label{fig-PitSubM05_MM206_2}
	\end{subfigure}	}
	\caption{Results concerning total pressure measures from test \textit{MM-0.5-B} with nozzle \textit{cMM05}. } 
	\label{fig-PitSubM05_MM206}
\end{figure}

Even though line-cavity systems are usually of $2^{nd}-$order, Figure \ref{fig-step} indicates that the total pressure line is so over-damped that its response closely resembles that of a $1^{st}-$order system.  
As such, its response is determined by the \textit{static sensitivity} $K$ and its \textit{time constant} $\tau$. Since $P_{line}=P_{tip}$ after the transient has expired, then it can be assumed that $K=1$. The time constant can be estimated by plotting quantity $Z(t)=\ln{\left(1-\frac{P_{line}(t)}{K P_{tip}(t)}\right)}$ as a function of time, as shown in Figure \ref{fig-tau}. The slope of the interpolating straight line is then related to the time constant: $\frac{d Z}{d t}=\frac{-1}{\tau}$. Data points fall very close to a straight line, confirming that the line response can be assumed of $1^{st}-$order type in the present case \citep{Doeblin1990}. The best fit line for all configurations was determined, giving the respective time constant values, as indicated in the legend, and allowing to quantify the different response speeds. 
The use of one \textit{Schaevitz} transducer more than doubles the response time with respect to the line-only case. Adding another sensor of the same type further increases it by a factor of $\sim 1.5$.
For comparison, one \textit{Kulite} transducer instead increases the time constant of the single line by only $\sim 50\%$.\\
Given these findings, testing with nozzle \textit{cMM05} was repeated by changing sensors type and also reducing their number to limit the overall volume on the Pitot tube total pressure line to the very minimum. Thus, the differential transducer devoted to direct probe kinetic head measurement was removed.
Also, the \textit{Schaevitz} sensor measuring the total pressure difference was substituted with a \textit{Kulite} one with a comparably low full scale of $\SI{0.7}{bar}$ and uncertainty of $\SI{1.7}{mbar}$.
These changes yield an estimated overall volume of the Pitot tube total pressure line of $\sim \SI{2700}{mm^3}$. This is less than a third of the previous configuration with the two \textit{Schaevitz} sensors, which had an overall volume of $\sim \SI{9000}{mm^3}$.
These volumes were estimated considering lines length, tubing internal diameter, fittings internal passages and transducers internal chambers as declared by components manufacturers or as directly measured.\\
The \textit{Schaevitz} differential transducer measuring $\Delta P_{ts,line}$ on the line exiting the plenum was also replaced by a \textit{Kulite} sensor with $\SI{5.9}{bar}$ full scale and an uncertainty of $\SI{10}{mbar}$.\\
Results concerning total pressure measures of tests with the reduced volume configuration are illustrated in Figure \ref{fig-PitSubM05_MM206}. Repeatability was verified and only one test is here reported (test \textit{MM-0.5-B}).
The system behavior has improved considerably, given that the Pitot tube total pressure is always superposed to $P_{t,ref}$ and $P_{t,line}$, as visible in Figure \ref{fig-PitSubM05_MM206_1}. This is confirmed by the total pressure difference $\Delta P_t$ in Figure \ref{fig-PitSubM05_MM206_2}, which is now always positive and is significantly lower in absolute terms, with a value of $\SI{6}{mbar}$ ($<1\%$ of the kinetic head) at the beginning and $\SI{1}{mbar}$ at the end of the test.\\
Thus, the pneumatic system can now be considered adequate for Pitot tube total pressure measures aswell, in the present configuration employing a probe with total pressure tap of $\SI{0.6}{mm}$ in diameter and subject to the \textit{TROVA} blow-down transient with a frequency content of the order of $\SI{1}{Hz}$.


\section{Conclusions and Outlook}
\label{sec-conclusions}

The present work reports the step-by-step development, testing and commissioning of a pneumatic system that will enable the calibration of probes operating with non-ideal flows representative of ORC working conditions.
A pneumatic lines scheme involving nitrogen flushing was purposely implemented to allow absolute and differential pressure measurements with probes in the test section of the Test Rig for Organic VApors blow-down wind tunnel at Politecnico di Milano, whilst avoiding issues linked to possible condensation such as the mass sink phenomenon. 
Commissioning of the complete system was performed with testing of an L-shaped Pitot tube in non-ideal subsonic flows of siloxane MM vapor at Mach numbers $M=0.2$ and $0.5$. 
Measurement delay issues were highlighted thanks to direct comparison with plant reference quantities, were verified through a dynamic testing procedure, and were solved by reducing the overall pneumatic lines volume by employing differential pressure transducers featuring a smaller internal cavity. \\
This work sets the foundations for future pressure probes calibration and for their use for characterization of non-ideal subsonic and supersonic flows of organic vapors, such as in the direct evaluation of shock total pressure losses and the testing of ORC turbine blade cascades.
Moreover, the paper presents detailed descriptions of the various development steps and procedures. These can be useful for anyone wishing to develop a pneumatic system for pressure probes testing in similar conditions, even beyond the ORC field and in any type of transient flow subject to line condensation.

\section{Acknowledgments}
The present research was funded by ERC Proof of Concept Grant N. 875015, project \textit{PROVA: Pitot PRobe for non-ideal compressible flows of organic VApors for renewable energy applications} under the Horizon 2020 Framework Programme, call: ERC-2019-PoC. \\
The authors wish to thank Ing. Gioele De Donati for his contribution to the experimental campaigns here reported.


\bibliography{Biblio_Measurement.bib}

\end{document}